\documentclass[twocolumn]{aastex631}

\shorttitle{A metallicity study of F, G, K and M dwarfs in the Coma Berenices}
\shortauthors{Souto et al.}
\graphicspath{{./}{figures/}}

\begin{document}

\title{A metallicity study of F, G, K and M dwarfs in the Coma Berenices open cluster from the APOGEE survey}

\correspondingauthor{Diogo Souto}
\email{diogosouto@academico.ufs.br}

\author[0000-0002-7883-5425]{Diogo Souto}
\affiliation{Departamento de F\'isica, Universidade Federal de Sergipe, Av. Marechal Rondon, S/N, 49000-000 S\~ao Crist\'ov\~ao, SE, Brazil}

\author[0000-0001-6476-0576]{Katia Cunha}
\affiliation{Observat\'orio Nacional/MCTIC, R. Gen. Jos\'e Cristino, 77,  20921-400, Rio de Janeiro, Brazil}
\affiliation{Steward Observatory, University of Arizona, 933 North Cherry Avenue, Tucson, AZ 85721-0065, USA}

\author{Verne V. Smith}
\affiliation{NSF's NOIRLab, 950 North Cherry Avenue, Tucson, AZ 85719, USA}

\begin{abstract}

We present a study of metallicities in a sample of main sequence stars with spectral types M, K, G and F ($T_{\rm eff}$ $\sim$ 3200 -- 6500K and log $g$ $\sim$ 4.3 -- 5.0 dex) belonging to the solar neighborhood young open cluster Coma Berenices. 
Metallicities were determined using the high-resolution (R=$\lambda$/$\Delta$ $\lambda$ $\sim$ 22,500) NIR spectra ($\lambda$1.51 -- $\lambda$1.69 $\mu$m) of the SDSS-IV APOGEE survey. 
Membership to the cluster was confirmed using previous studies in the literature along with APOGEE radial velocities and Gaia DR2.
An LTE analysis using plane-parallel MARCS model atmospheres and the APOGEE DR16 line list was adopted to compute synthetic spectra and derive atmospheric parameters ($T_{\rm eff}$ and log $g$) for the M dwarfs and metallicities for the sample.
The derived metallicities are near solar and are homogeneous at the level of the expected uncertainties, in particular when considering stars from a given stellar class. The mean metallicity computed for the sample of G, K, and M dwarfs is $\langle$[Fe/H]$\rangle$ = +0.04 $\pm$ 0.02 dex; however, the metallicities of the F-type stars are slightly lower, by about 0.04 dex, when compared to cooler and less massive members. Models of atomic diffusion can explain this modest abundance dip for the F dwarfs, indicating that atomic diffusion operates in Coma Berenices stars. The [Fe/H] dip occurs in nearly the same effective temperature range as that found in previous analyses of the lithium and beryllium abundances in Coma Berenices. 

\end{abstract}


\section{Introduction} \label{sec:intro}

Open clusters and their stellar members are excellent laboratories to study the effects of stellar evolution on stellar abundances, as its members share the same distance, age, and initial chemical composition. Under the assumption that cluster stars were born at the same time and from the same molecular cloud, such constraints make open clusters good laboratories for conducting detailed chemical studies in order to, for example, investigate departures from chemical homogeneity that can be used to identify processes that operate in stellar atmospheres which can modify their surface abundances. 
Such processes include convective dredge-up mechanisms, which are observed as a product of stellar evolution in red giant stars \citep{Becker1979,Lagarde2012,Salaris2015}, or atomic diffusion, driven by both gravitational settling and radiative acceleration, in dwarf stars \citep[e.g.,][]{Chaboyer1995a,Richard2005,Dotter2017,Souto2018, Gao2018, BertelliMotta2018, Liu2019}.

The APOGEE survey observed a number of open clusters along with Galactic disk stellar populations \citep[see also, ][]{Donor2020}; among the solar neighborhood clusters is Coma Berenices (Melotte 111; Collinder 256; hereafter Coma Ber), which is one of the closest open clusters to the Sun.
Coma Ber (RA = 221.35280, and DEC = +84.02485 (J2000)) is a young open cluster with an estimated age of 600 -- 800 Myr \citep{Casewell2006,Kraus2007,Casewell2014,Tang2018comaber}. 
The estimated distance to Coma Ber is $\sim$ 85 $\pm$ 7.1  pc \citep{Tang2018comaber} and given its proximity to the Sun, this cluster suffers from low stellar reddening (E(B-V) $<$ 0.01) \citep{Nicolet1981,Taylor2007}.
Previous studies of Coma Ber in the literature, both based on photometry, and on low and high resolution spectroscopy have found it to have near solar metallicity  \citep{Nissen1981,Boesgaard1987,Cayrel1988,Boesgaard1989, FrielBoesgaard1992,Burkhart2000,Gebran2008,Terrien2014}. 

The APOGEE survey also observed the older and more distant solar metallicity open cluster M67 and \cite{Souto2018} used APOGEE spectra to study diffusion in samples of M67 stars in different evolutionary states, from the main sequence, to turnoff stars, and up the red giant branch. That study found abundance variations in M67 turnoff stars and demonstrated that atomic diffusion models could explain the observed abundance trends well. 
In a larger study of M67 chemical abundances, \citep{Souto2019a} were able to analyze dwarf stars as faint and cool (T$_{\rm eff}\sim$4850K) as spectral type K, but M67 M dwarfs were too faint to have been observed by APOGEE. \cite{Souto2017,Souto2020} have analyzed field M dwarfs observed as part of APOGEE and shown that detailed abundance distributions can be derived from APOGEE spectra.
The proximity of Coma Ber to the Sun (being some 10x closer than M67), on the other hand, offered the opportunity to expand APOGEE observations to cooler main-sequence stars in this open cluster, reaching the coolest M dwarfs ($T_{\rm eff}$ $\sim$ 3100 K).  Coma Ber M dwarf abundances are  valuable in verifying the APOGEE metallicity scale for the M dwarfs (whose spectra have a large contribution from water lines) and whether there is an agreement with the results from the warmer K-type stars. Moreover, the APOGEE observations of Coma Ber targeted the hotter main-sequence F stars (Beaton et al. 2021 submitted) and such a data set provides a wider T$_{\rm eff}$ range along the main sequence in which to probe diffusion in a much younger open cluster. 

In this study, we will use the APOGEE spectra to first verify membership to the Coma Ber cluster by combining Gaia DR2 \citep{GaiaCollaboration2018_HRdiag} parallaxes and proper motions, with APOGEE radial velocities (RV). We will present a detailed spectroscopic analysis of a selected sample of  F, G, K, and M dwarf members of Coma Ber and derive their stellar parameters, effective temperatures, surface gravities, and metallicities. 
Such a data set, covering an extended $T_{\rm eff}$-space, will allow us to probe the level of internal precision of the metallicity results obtained from our methodology and to search for possible systematic differences that may occur in different $T_{\rm eff}$ regimes due to atomic diffusion processes.
In addition, the results in this study will be valuable to compare and verify the parameters and metallicities publicly available in the APOGEE DR16, derived using the automated abundance pipeline ASPCAP \citep{GarciaPerez2016}, which has not been optimized for the study of cool dwarfs given that the APOGEE survey is primarily a survey of red-giant stars.
This paper is organized as follows: in Section 2, we describe the observational data and the membership selection; Section 3 presents the abundance analysis and the atmospheric parameters, and in Sections 4 and 5, we present the main results and conclusions of this work, respectively.

\section{OBSERVATIONS AND SAMPLE}

\begin{figure*}
	\includegraphics[width=0.49\linewidth]{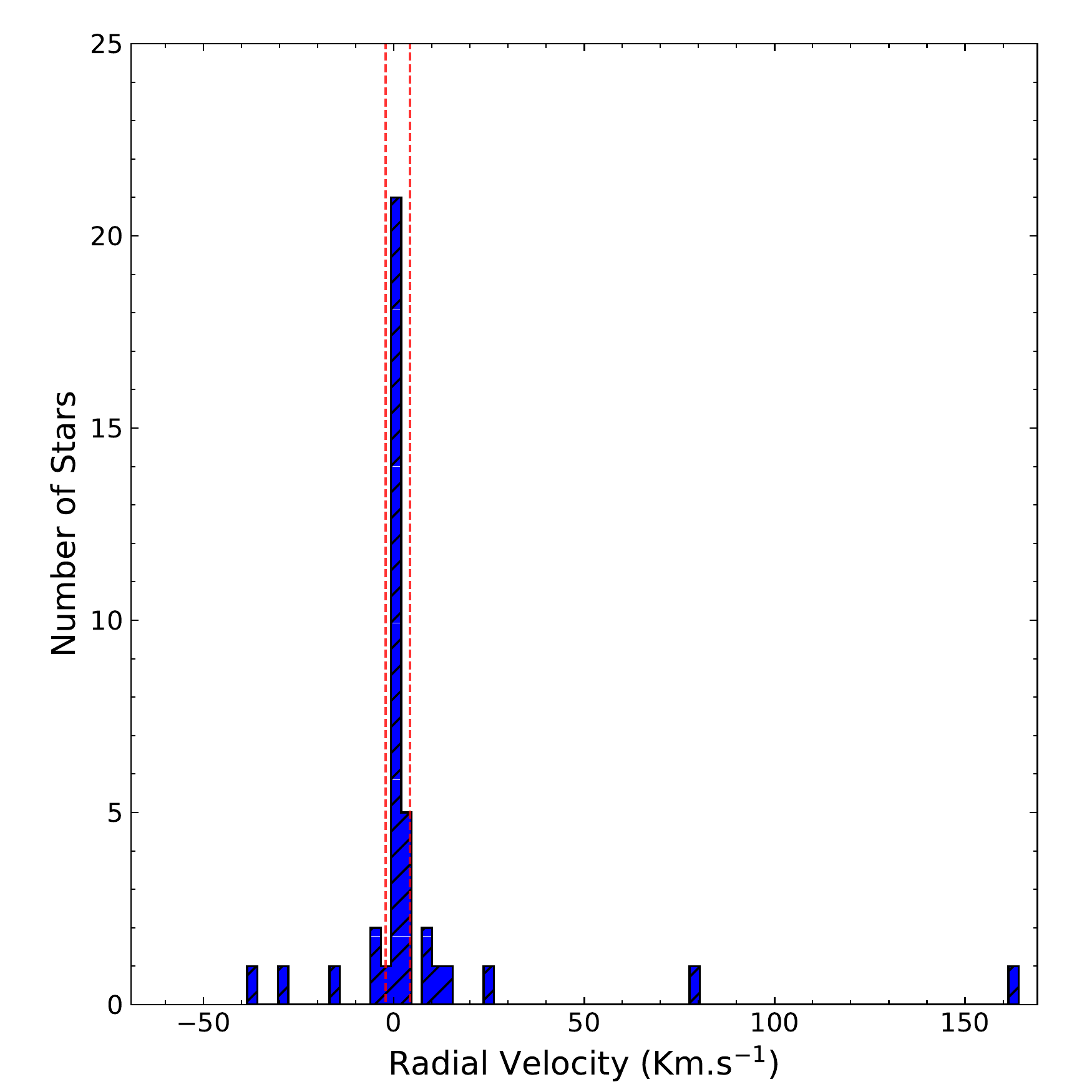}
	\includegraphics[width=0.49\linewidth]{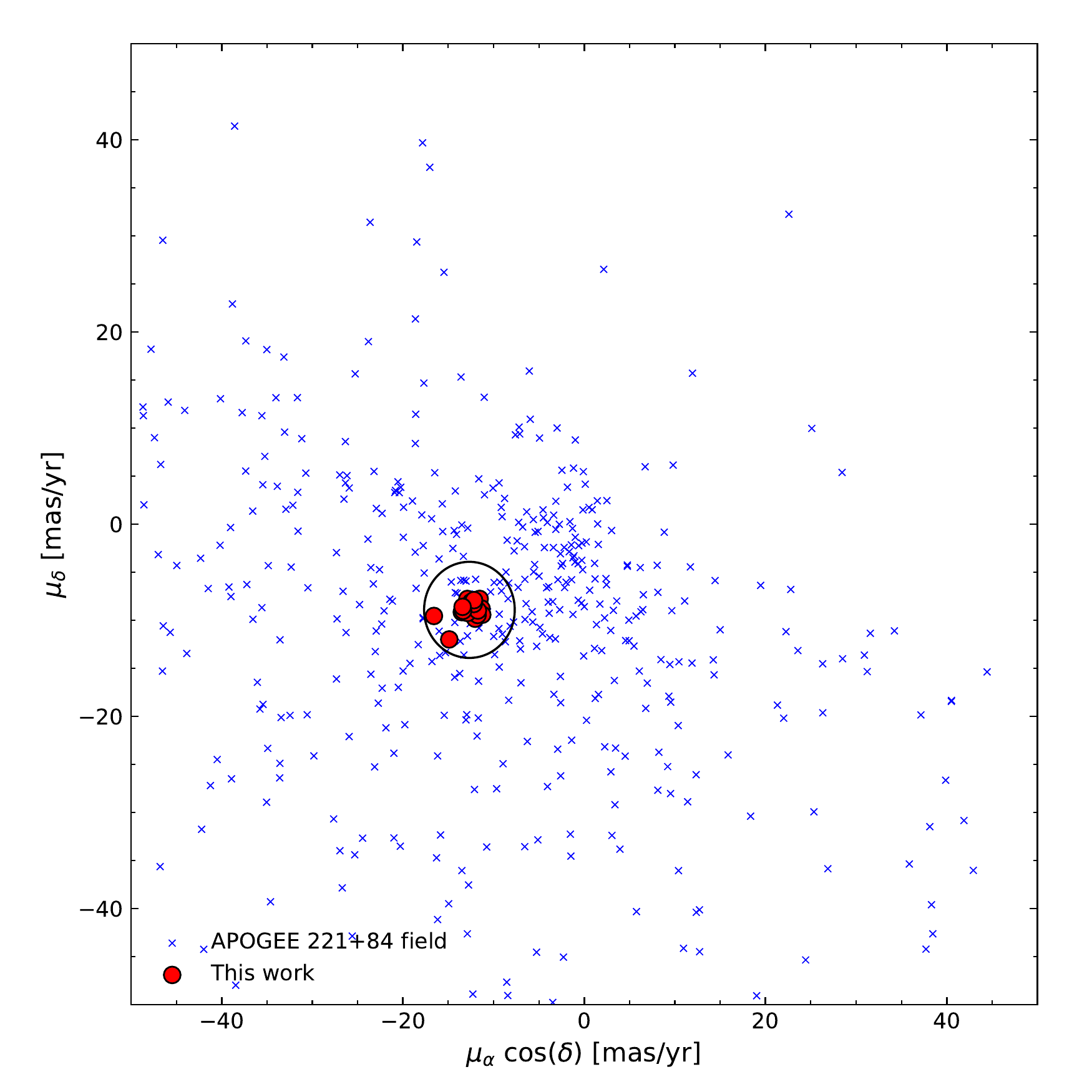}	
	\includegraphics[width=0.49\linewidth]{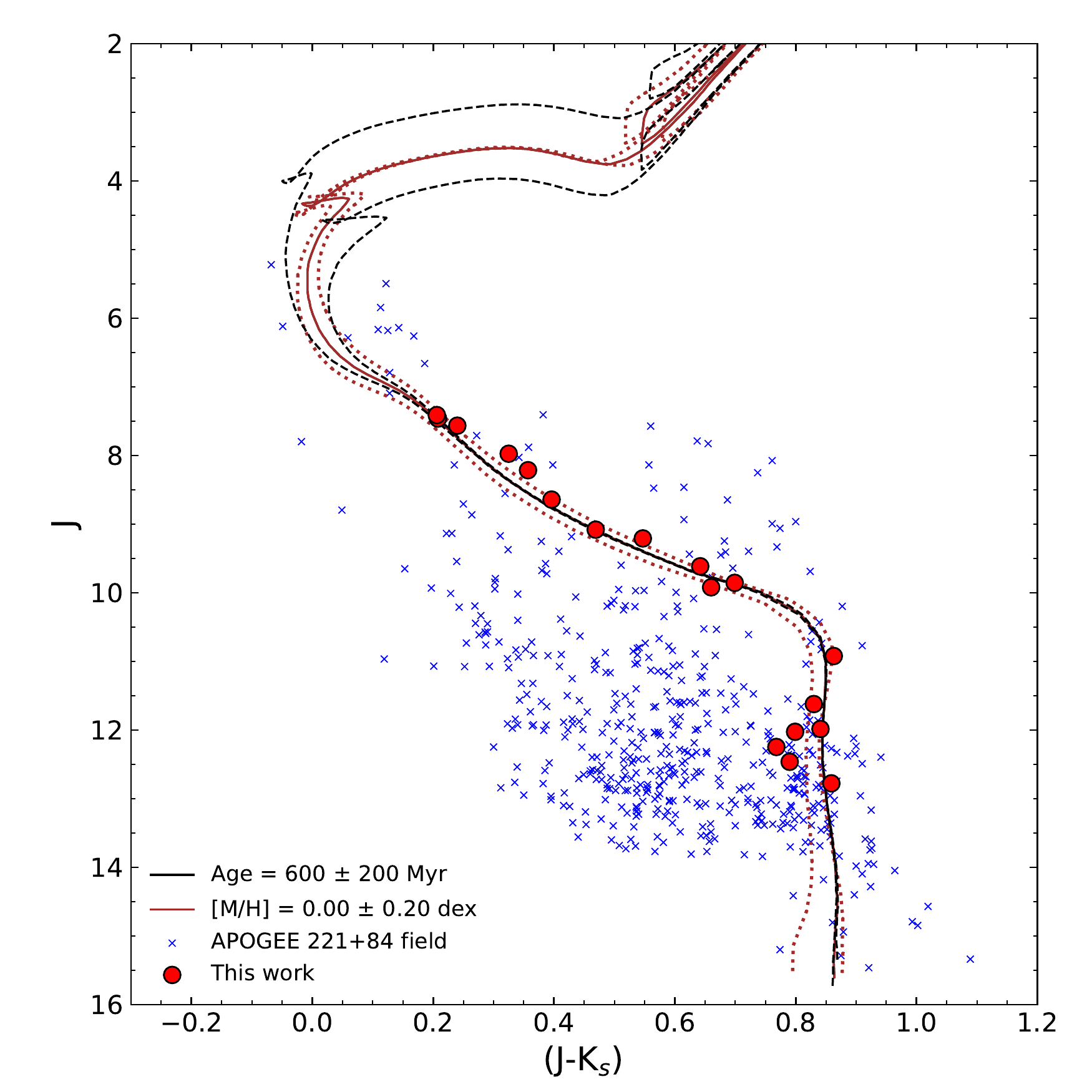}
    \includegraphics[width=0.49\linewidth]{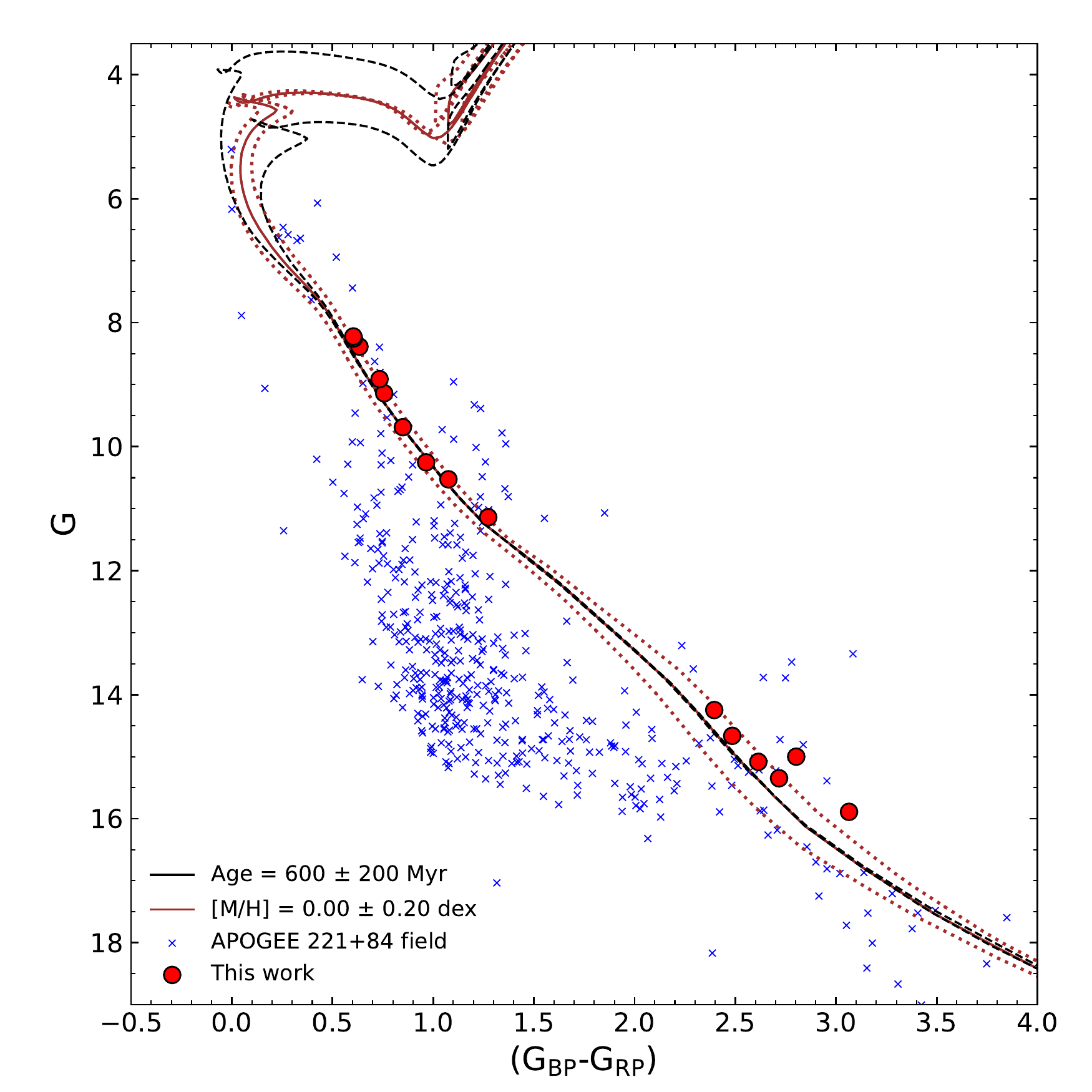}
    \caption{Upper left panel shows the radial velocity distribution from the DR16 APOGEE radial velocity pipeline for the observed APOGEE targets in common with the Coma Ber membership study of \cite{Tang2018comaber} study. The red dashed lines indicate the RV range determined for the cluster. Upper Right panel shows the Gaia DR2 proper motions for the target stars. Lower right and left panels show the ($J$ - $K_{s}$) versus $J$, and (G$_{\rm BP}$ - G$_{\rm RP}$) versus $G$ diagrams using 2MASS and Gaia DR2 photometry, respectively. MIST isochrones are also shown as the brown and black lines. The blue crosses correspond to all APOGEE targets observed in the APOGEE 221+84 field.}
    \label{membership1}
\end{figure*}

\subsection{The APOGEE Spectra}

The spectra analyzed in this study were obtained by the SDSS-III and SDSS-IV APOGEE  \citep{Eisenstein2011,Blanton2017} surveys \citep{Majewski2017}.
APOGEE operates two cryogenic, multi-fiber (300 fibers) spectrographs  \citep{Wilson2010,Gunn2006} with spectral coverage in the NIR (H-band) between $\lambda$1.51 $\mu$m - $\lambda$1.69 $\mu$m. The APOGEE North and South spectrographs are currently located in both hemispheres, on the 2.5m telescopes at APO (Apache Point Observatory, New Mexico, USA) and at LCO (Las Campanas Observatory, La Serena, Chile). 

APOGEE is a dedicated survey to study Galactic archaeology via the chemical abundance analysis of high-resolution spectra of red giant stars. Although the primary targets for the APOGEE survey are Galactic red giants, APOGEE has also observed main-sequence stars  (Beaton et al. 2021 submitted; \cite{Zasowski2017}), and in particular, those which are important for calibration purposes, e.g., dwarfs members of open clusters, dwarfs having astereoseismology data observed by the Kepler mission \citep{Pinsonneault2018}, as well as, hot main-sequence stars that serve as telluric calibrators for APOGEE observations.

The reduced spectra analyzed in this study were obtained from the publicly available 16-th APOGEE data release \citep[DR16;][]{Holtzman2018,Jonsson2018}. 
Given its proximity to the Sun, the Coma Ber open cluster encompasses about 7 degrees in diameter in the sky. The general region of the Coma Ber open cluster was targeted by the APOGEE main survey with observations of the APOGEE 221+84 field.
There were several APOGEE visits for this field so that the required minimum signal-to-noise of 100 could be reached for the fainter dwarf targets (M dwarfs), while for brighter dwarfs, the S/N reached was extremely high, in excess of $\sim$ 2000 (Table \ref{parameters}). 
Each APOGEE visit produces a radial velocity measurement, and given the number of visits, the internal uncertainty in the radial velocity measurements (sigma RV) is small, reaching 0.45 km s$^{-1}$. We used the DR16 radial velocities, and the radial velocity scatter derived by the APOGEE RV pipeline \citep{Nidever2015} to further confirm membership and identify possible binary stars in the selected sample.

\subsection{Membership Analysis}

We used the membership study of \cite{Tang2018comaber} as a reference to select stellar members of the Coma Ber open cluster observed by APOGEE. 
\cite{Tang2018comaber} identified 148 stars as stellar and sub-stellar members of Coma Ber based on their photometry, proper motions, and distances using the databases of 2MASS \citep{2MASS}, UKIDSS \citep{UKIDSS}, URAT1 \citep{URAT1}, and Gaia/DR2 \citep{GaiaCollaborationDR2}, when available. However, \cite{Tang2018comaber} did not use radial velocities to constrain membership to the cluster, which can be used here.

We cross-matched the target list of 148 bona fide members from \cite{Tang2018comaber} with the sample of 563 stars observed in the APOGEE 221+84 field; this found 39 stars in common, all of them having Gaia DR2 data available. 
We used the DR16 radial velocities, and the radial velocity scatter derived by the APOGEE RV pipeline \citep{Nidever2015} to further confirm membership and identify possible binary stars in the selected sample.
Figure \ref{membership1} (top left panel) shows the distribution of radial velocities for the 39 selected targets. The RV distribution obtained for this sample is peaked around 1.00 km s$^{-1}$ but shows easily identifiable outliers; the median RV and standard deviation of the median are $<$RV$>$ = 1.05 $\pm$ 1.59 km s$^{-1}$.
To remove possible non-members, we adopted the median RV as representative of the open cluster members and applied a 2-$\sigma$ RV cut to the sample and to remove possible binary stars we used a cut at 1.00 km s$^{-1}$ in RV scatter \citep[see][]{El-Badry2018} for a discussion on the RV scatter threshold for APOGEE observations). 
We also inspected the APOGEE spectra of the selected targets, and we note that two M dwarfs (2M12260848+2439315 and 2M12214070+2707510) are spectroscopic binaries (these M dwarfs had also been previously identified as SB2 in the APOGEE study by \citep{Skinner2018}),  and one F dwarf (2M12273836+2554435) was observed as a telluric standard; these stars were not analyzed here.
The sample M dwarf 2M12193796+2634445 was also identified by \cite{Skinner2018} as a likely spectroscopic binary star. However, the authors concluded that this star system did not exhibit a radial velocity separation sufficiently large enough to measure its RVs from their RV extraction method. 
The selected members of Coma Ber to be analyzed in this study are presented in Table \ref{parameters}. (We note that all of these, except one (2M12241121+2653166), are also considered to be members of Coma Ber with membership probabilities higher than 87\% in \cite{Kraus2007}.)

Based on the APOGEE RV data for the selected sample of Coma Ber members, the mean cluster RV is $<$RV$>$ = 0.86 $\pm$ 0.72 ($\pm$0.47) km s$^{-1}$. This mean RV for the cluster obtained is slightly higher than the Gaia DR2 value for Coma Ber reported by \cite{GaiaCollaboration2018_HRdiag} of $<$RV$>$ = 0.21 $\pm$ 0.13 km s$^{-1}$, although when considering the respective uncertainties, the mean radial velocities are in marginal agreement.
In the top right panel of Figure \ref{membership1} we show the proper motions from Gaia DR2 for the target sample. The mean cluster value according to Gaia DR2 is $\mu_{\alpha}$ cos($\delta$) = -12.111 mas/yr, and $\mu_{\delta}$ = -8.996 mas/yr \citep{GaiaCollaborationDR2}, and we use this value as a reference in  Figure \ref{membership1}; we also show a circle corresponding to a radius of $\pm$ 3.0 mas/yr accounting for uncertainties.

\begin{deluxetable*}{lccccccccccccc}
\tablecaption{Sample Stars \& Stellar Parameters\label{parameters}}
\tablewidth{700pt}
\tabletypesize{\scriptsize}
\tablehead{
2MASS ID &J &J-$K_{s}$ &G &G$_{\rm BP}$-G$_{\rm RP}$ & Visits	& RV & SNR  &   d(pc) BJ18$^{\dagger}$    & $v sin (i)$   &$T_{\rm eff}$	& log $g$  & [Fe/H]	& $\sigma$([Fe/H])}
\startdata
2M12221448+2526563$^{*}$    	&13.980		&0.900	&17.209	&3.279	&7	&0.52$\pm$0.50  &40 	&87.14$\pm$1.40	& --	& --	& --	& --	& --\\
2M12193796+2634445		&12.776	&0.859	&15.890	&3.065	&8	&2.92$\pm$0.79	&147	&87.84$\pm$2.43	&40.0	&3110	&5.00	&0.05	&0.09\\
2M12264027+2718434		&12.462	&0.79	&15.349	&2.717	&11	&1.39$\pm$0.21	&196	&81.57$\pm$0.53	&10.5	&3314	&4.98	&0.07	&0.09\\
2M12201448+2526072		&12.246		&0.768	&15.084	&2.614	&10	&1.80$\pm$0.15	&186	&92.11$\pm$0.61	&$<$7.0	&3373	&4.83	&0.03	&0.09\\
2M12231356+2602185		&12.025		&0.799	&15.000	&2.802	&11	&1.46$\pm$0.48	&136	&81.56$\pm$1.32	&11.0	&3279	&4.99	&0.08   &0.09\\
2M12255421+2651387		&11.984		&0.841	&14.664	&2.485	&11	&1.14$\pm$0.22	&290	&87.46$\pm$0.43	&$<$7.0	&3429	&4.88	&-0.01	&0.09\\
2M12250262+2642382		&11.621	&0.83	&14.248	&2.396	&11	&0.61$\pm$0.11	&385	&83.79$\pm$0.40	&$<$7.0	&3474	&4.81	&0.01	&0.09\\
2M12241121+2653166		&10.921		&0.863	&13.133	&	&11	&0.63$\pm$0.15	&507	&86.13$\pm$0.26	&$<$7.0	&3780	&4.70	&0.04	&0.13\\
2M12232820+2553400		&9.92		&0.66	&11.687	&	&11	&0.33$\pm$0.10	&401	&86.02$\pm$0.33	&$<$7.0	&4460	&4.64	&0.05	&0.03\\
2M12265103+2616018		&9.855		&0.699	&11.538	&	&11	&0.33$\pm$0.10	&919	&85.76$\pm$0.43	&$<$7.0	&4508	&4.65	&0.02	&0.03\\
2M12211561+2609140		&9.614		&0.642	&11.139	&1.273	&11	&0.45$\pm$0.15	&627	&84.71$\pm$0.32	&$<$7.0	&4717	&4.66	&0.04	&0.03\\
2M12285643+2632573		&9.208		&0.547	&10.526	&1.076	&11	&1.06$\pm$0.05	&1052	&84.15$\pm$0.34	&$<$7.0	&5173	&4.62	&0.03	&0.03\\
2M12240572+2607430		&9.08		&0.469	&10.253	&0.965	&8	&0.36$\pm$0.29	&933	&88.67$\pm$0.28	&$<$7.0	&5406	&4.69	&0.03	&0.04\\
2M12270627+2650445		&8.642		&0.396	&9.686	&0.850	&8	&0.39$\pm$0.12	&1223	&89.35$\pm$0.83	&$<$7.0	&5701	&4.58	&0.02	&0.04\\
2M12214901+2632568		&8.214		&0.357	&9.139	&0.757	&8	&0.64$\pm$0.17	&1406	&86.29$\pm$0.36	&$<$8.5	&5949	&4.50	&0.02	&0.04\\
2M12204557+2545572		&7.974		&0.325	&8.911	&0.734	&11	&-0.36$\pm$0.42	&949	&84.24$\pm$0.43	&$<$7.0	&5899	&4.33	&0.04	&0.04\\
2M12215616+2718342		&7.565		&0.24	&8.387	&0.634	&14	&-0.03$\pm$0.18	&2080	&85.14$\pm$0.50	&20.0	&6480	&4.33	&-0.01	&0.04\\
2M12234101+2658478		&7.461		&0.208	&8.253	&0.607	&14	&0.90$\pm$0.44	&2417	&87.24$\pm$0.53	&25.0	&6478	&4.28	&0.01	&0.04\\
2M12255195+2646359		&7.411		&0.206	&8.222	&0.604	&14	&1.47$\pm$0.41	&2571	&86.10$\pm$0.38	&15.0	&6530	&4.29	&0.00	&0.04\\
\enddata
\tablecomments{$^{*}$ Not analysed due to low signal to noise ratio. $^{\dagger}$ Distances from \cite{Bailer-Jones2018} (BJ18).}
\end{deluxetable*}

Using some of the same early APOGEE data \citep[from DR10; ][]{DR10} in conjunction with IRTF - SpeX medium-resolution spectra, and photometry from the Kilodegree Extremely Little Telescope (KELT), \cite{Terrien2014} studied membership in Coma Ber and confirmed at the time six new M dwarfs as members. 
The membership analysis discussed here, which now considers Gaia DR2 data, added three additional M dwarf members of Coma Ber (2M12193796+2634445, 2M12221448+2526563 and 2M12260848+2439315)
and confirmed the membership for ten K and M dwarfs from \cite{Terrien2014}.

\begin{figure*}
	\includegraphics[width=1\linewidth]{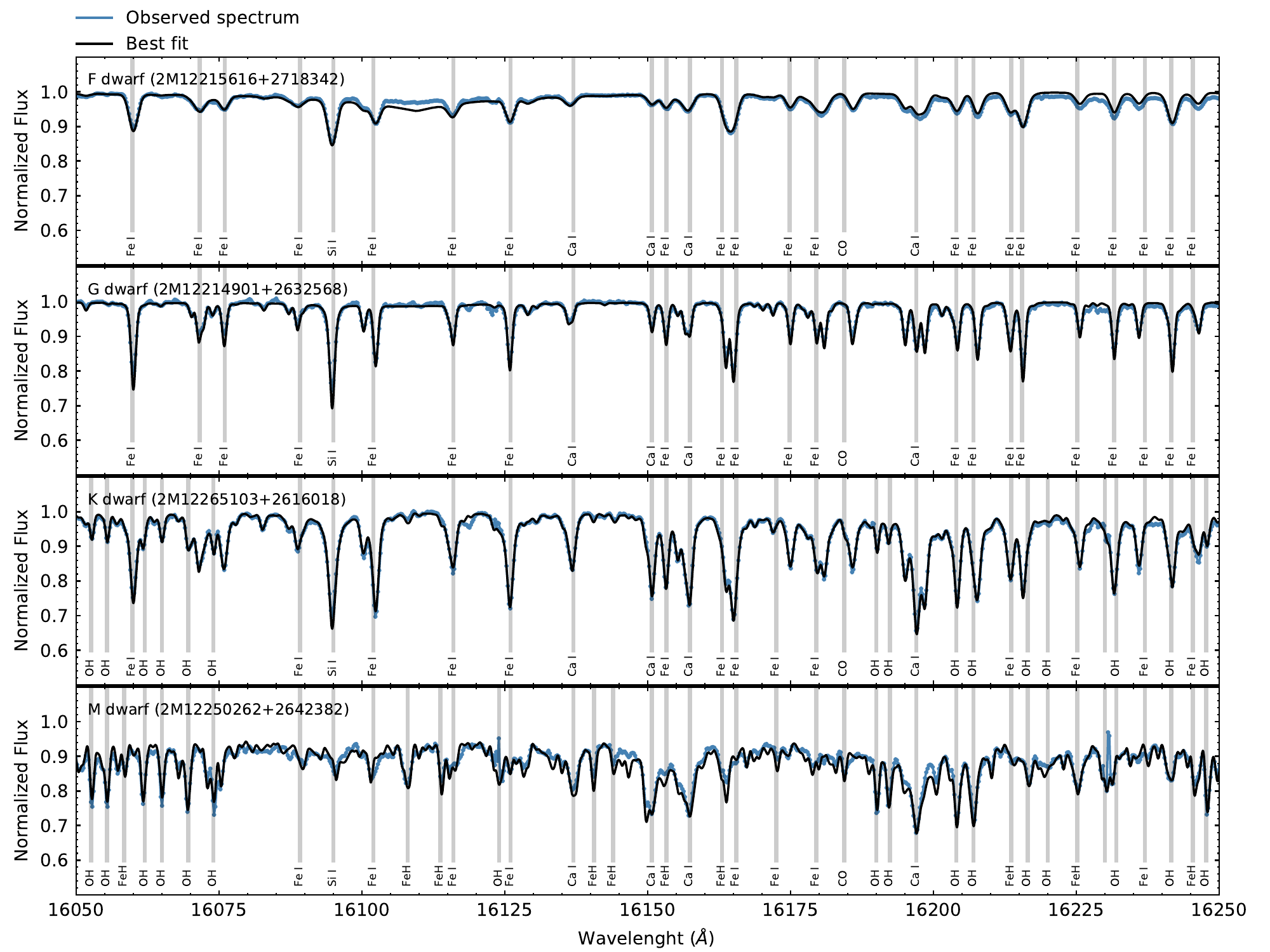}
    \caption{Portions of the APOGEE spectra of four target stars roughly covering the range in effective temperature of the target stars. This illustrates the sensitivity of spectral features to the effective temperature from an F dwarf ($T_{\rm eff}$ = 6480 K, top panel) to an M dwarf ($T_{\rm eff}$ = 3437 K, bottom panel).}
    \label{spectra}
\end{figure*}

In the lower panels of Figure \ref{membership1} we show color-magnitude diagrams (CMDs) from 2MASS \citep{2MASS} ($J$-$K_{S}$ -- $J$) and Gaia DR2 ($G_{\rm BP}$-$G_{\rm RP}$ -- $G$); the studied targets are shown as red circles and the sample of APOGEE targets observed in the 221+84 field are shown as blue crosses. We also show, for comparison, six sets of MIST isochrones \citep{Dotter2016,Choi2016} encompassing the age and metallicity of Coma Ber from previous studies: a solar metallicity 600 Myr isochrone (shown as solid brown) along with isochrones $\pm$ 0.20 dex in metallicity (shown as brown dashed lines); the dashed black lined isochrones correspond to ages of 400 and 800 Myr and solar metallicity.
The studied stars from Coma Ber (red circles) are clearly not evolved, not having reached the turnoff point. Overall, their color and magnitudes put them close to the displayed MIST isochrones, with the exception of two slightly more discrepant points (M dwarfs) both in the $J$-$K_{S}$--$J$ and $G_{\rm BP}$-$G_{\rm RP}$ --$G$  diagrams. The locus of most of the targets from the APOGEE field falls quite far from the isochrones, leaving little doubt about their non-pertinence to the cluster.

\section{Fe Abundance analysis}

The abundance analysis adopted in this work basically consists of generating spectral syntheses and finding the best matches to the observations. 
To generate the synthetic spectra, we use 1-D plane-parallel local thermodynamic equilibrium (LTE) MARCS model atmospheres \citep{Gustafsson2008} and the Turbospectrum code \citep{AlvarezPLez1998,Plez2012}.
We adopted the most recent APOGEE line list used in DR16, which includes the transitions needed for the analysis of the H band spectra of M dwarfs, such as H$_{2}$O lines \citep{Barber2006} and FeH transitions \citep{Hargreaves2010}. The methodology adopted to construct the DR16 line list is described in detail in \cite{Smith2021}. 

Individual Fe abundances (from Fe I and FeH lines for the M dwarfs and Fe I lines for the F, G, K dwarfs), were determined using the semi-automatic mode of the BACCHUS wrapper \citep{Masseron2016}, with which we can manually adjust the pseudo-continuum, lambda displacements, and the line broadening, whenever needed. We adopted solar C, N, O for the analysis of the F, G, K dwarfs. 
We derived the abundances of C (from CO lines) and O (from H$_{2}$O and OH lines) for the M dwarfs, while the nitrogen abundances were kept solar.

\subsection{Atmospheric Parameters}

As part of a 1-D plane parallel analysis, we determined the microturbulent velocity parameter ($\xi$), which was varied with the goal of obtaining the minimum spread of the Fe I line abundances over different values of $\xi$. For all M dwarfs, the adopted microturbulent velocities were $\sim$ 1 km s$^{-1}$. In previous studies we tested that these were adequate microturbulent velocity values to fit the spectra of M dwarfs, noting that M dwarf parameters are not very sensitive to the microtubulent velocity. 
For the K and G dwarfs similar microturbulent velocity values were obtained.
For the F dwarfs, higher values of microturbulent velocities were obtained of $\sim$ 1.80 km s$^{-1}$. The sensitivity of the derived metallicities to the microturbulent velocity parameter is in Table \ref{uncert}. 
For all syntheses, we adopted a Full-Width Half-Maximum (FHWM) of 730 m\AA{}, which corresponds to the APOGEE instrumental resolution broadening.
Three M, one G, and three F dwarfs in our sample have measurable \textit{v sin i} beyond the APOGEE resolution threshold of $\sim$ 7 km s$^{-1}$ (Table \ref{parameters}).

The derivation of stellar parameters for the studied stars followed different methodologies for the M dwarfs when compared to the F, G, and K dwarfs, given that their APOGEE spectra show distinct spectral characteristics. In particular, the APOGEE spectra of M dwarfs have significant contributions from water and FeH lines, which are not present in the warmer stars.

The methodology adopted for the determination of $T_{\rm eff}$ and log $g$ for the M dwarfs consisted in finding the best model atmosphere that brought the oxygen abundances from OH and H$_{2}$O lines into an agreement, by searching for a consistent solution for the pairs $T_{\rm eff}$--A(O) and log $g$--A(O). 
For the F, G, and K dwarfs, we adopted the effective temperatures directly from the APOGEE DR16 (raw non-calibrated values), which were derived using the ASPCAP pipeline  \citep{GarciaPerez2016} from overall fits of synthetic models computed with the same radiative transfer code (Turbospectrum) and the same DR16 APOGEE line list used here for the computation of synthetic spectra.
The surface gravity values for the F, G, and K targets were determined using the fundamental relation presented in Equation 1, which is derived directly from the Stefan–Boltzmann law.

\begin{center}
\begin{equation}
\log{g} = \log{g_{\odot}} + \log\left(\frac{M_{\star}}{M_{\odot}}\right) +
4\log\left(\frac{T_{\star}}{T_{\odot}}\right) + 0.4(M_{bol,\star} - M_{bol,\odot})
\end{equation}
\end{center} where stellar masses and bolometric magnitudes were determined from matching the MIST isochrones and assuming: [Fe/H] = 0.00, age = 600 Myr, E(B-V)=0.00, and distance modulus $\mu$ = 4.69 \citep{GaiaCollaboration2018_HRdiag,Tang2018comaber}.
The adopted solar values were: log $g_{\odot}$ = 4.438 dex, $T_{\rm eff, \odot}$ = 5772 K and $M_{bol,\odot}$ = 4.75, following the IAU recommendations in \cite{IAU_solar}.

Illustrations of the overall quality of the model fits to the observed APOGEE spectra are presented in Figure \ref{spectra}.
This figure shows the observed APOGEE spectra (in cyan) of four target stars in a selected spectral region between $\lambda$=16,050 -- 16,250 \AA.
The respective best fitting synthetic spectra for each star are shown in black. The targets in this figure were selected to illustrate the four $T_{\rm eff}$ regimes covered by the sample stars: the F dwarf 2M12215616+2718342 (top panel), the G dwarf 2M12214901+2632568 (second panel), the K dwarf 2M12265103+2616018 (third panel), and the M dwarf 2M12250262+2642382 (bottom panel). We can see that there are changes in spectral features as the effective temperature decreases from top to bottom. It is notable, for example, that the depths of the lines change significantly as a function of $T_{\rm eff}$, going through a maximum depth for the K dwarf; for example, the atomic Fe I and Si I lines reach maximum depths around 4500 K and decrease intensity for both higher and lower $T_{\rm effs}$.  
We can also see that the molecular transitions of OH and FeH are stronger for the M dwarf and that its spectrum has a depressed 'pseudo continuum', lowered by roughly 10 percent, which is due to absorption by a large number of water transitions throughout the spectrum.

\subsection{Estimated Uncertainties}

The uncertainties in the derived stellar parameters are estimated to be:
$\Delta$$T_{\rm eff}$ = $\pm$ 100 K, $\Delta$log $g$ = 0.20 dex and $\Delta$$\xi$ = $\pm$ 0.20 km s$^{-1}$ for the M dwarfs, and $\Delta$$T_{\rm eff}$ = $\pm$ 50K; $\Delta$log $g$ = $\pm$0.10 dex; $\Delta$$\xi$ = $\pm$ 0.20 km s$^{-1}$ for the F, G, K dwarfs \citep{Souto2018,Souto2020}. 
The uncertainties in the iron abundances derived for five effective temperature regimes representative of our target sample were computed from the abundance sensitivities to the errors in the atmospheric parameters ($\Delta$$T_{\rm eff}$, $\Delta$log $g$, and $\Delta$$\xi$) and model atmosphere metallicity ($\Delta$[M/H]). We also investigated the abundance deviations that would result from small uncertainties in the definition of the pseudo-continuum  ($\Delta$ Pseudo-continuum) for each representative star.
In Table \ref{uncert} we present the estimated uncertainties in the metallicities; these were obtained from the quadrature sum of $\Delta$A(Fe) corresponding to the adopted errors in the parameters. 
In summary, the uncertainties in the derived metallicities are small and $\sim$ 0.03 -- 0.04 dex for the F, G, K dwarfs, while these are found to be higher for the M dwarfs (0.09 and 0.13 dex for the late M- and early M- dwarfs, respectively). The dominant sources of uncertainties for the M dwarfs with $T_{\rm eff}$ $\sim$ 3800 is the higher abundance sensitivities to changes in $T_{\rm eff}$ and log $g$.

\begin{deluxetable*}{lccccccr}
\tablecaption{Fe Abundance Sensitivities due to Uncertainties in the Atmospheric Parameters\label{uncert}}
\tablewidth{700pt}
\tabletypesize{\scriptsize}
\tablehead{
Model Atmosphere   &   $\Delta$ $T_{\rm eff}$   &   $\Delta$ log $g$ &   $\Delta$ $\xi$   &   $\Delta$ [M/H]   & $\Delta$ Pseudo-continuum &    $\sigma$\\
 Parameters   &   (+50 / 100 K) & (+0.20 dex)   &   (+0.20 km s$^{-1}$) &   (+0.10 dex)   & &}
\startdata
F dwarf \\
$T_{\rm eff}$ = 6300 K; log $g$ = 4.30 dex   &   +0.03	&-0.01	&-0.01	&+0.00	&+0.02	&0.039 \\
G dwarf \\
$T_{\rm eff}$ = 5772 K; log $g$ = 4.44 dex   &   +0.03	&-0.02	&+0.00	&+0.01	&+0.02	&0.042 \\
K dwarf \\
$T_{\rm eff}$ = 4700 K; log $g$ = 4.60 dex   &   +0.00	&+0.02	&-0.01	&+0.01	&+0.02	&0.032 \\
Early M dwarf \\
$T_{\rm eff}$ = 3800 K; log $g$ = 4.80 dex   &   -0.09$^{*}$	&+0.09	&+0.00	&+0.01	&+0.02	&0.129 \\
Mid M dwarf \\
$T_{\rm eff}$ = 3300 K; log $g$ = 5.00 dex   &   +0.05$^{*}$	&-0.02	&-0.02	&+0.07	&+0.02	&0.093
\enddata
\tablecomments{$^{*}$ A larger $T_{\rm eff}$ uncertainty of 100 K was adopted for the M dwarfs. The $\sigma$ represents the sum in quadrature of the uncertainties.}
\end{deluxetable*}

\section{Results and Discussion}

The derived stellar parameters and metallicities for 18 Coma Ber stars are presented in Table \ref{parameters}, where the Solar iron abundance from \cite{Asplund2009} (A(Fe) = 7.45) was adopted as a reference.
We find that the Coma Ber open cluster stars have near solar metallicity with a small internal spread in the metallicities. 
In particular, if we split the targets in terms of $T_{\rm eff}$ regime, we obtain the following mean metallicities: 
$\langle$[Fe/H]$\rangle$ = 0.04 $\pm$ 0.03 dex for the seven M dwarfs, 
$\langle$[Fe/H]$\rangle$ = 0.04 $\pm$ 0.01 dex for the four K dwarfs, 
$\langle$[Fe/H]$\rangle$ = 0.03 $\pm$ 0.01 dex for the four G dwarfs, 
and $\langle$[Fe/H]$\rangle$ = 0.00 $\pm$ 0.01 dex for the three F dwarfs. 
Such results suggest that there is chemical homogeneity within each stellar class in Coma Ber stars. 
However, it is noted that the mean metallicity of the F stars is slightly lower than the ones from the G, K, and M, stars and this will be discussed in Section 4.2.

Table \ref{M_abu} presents the detailed line-by-line inventory of the individual Fe I and FeH line abundance measurements and the respective mean Fe abundance values for the studied M dwarfs, and Table \ref{FGK_abu} contains the line-by-line iron abundances from the Fe I lines measured for the FGK dwarfs.  The adopted metallicities for the M dwarfs in this study are the mean of Fe I and FeH abundances, when available, keeping in mind that the coolest M dwarfs ($T_{\rm eff}$ $<$ 3200 K) do not have Fe I lines strong enough to be measured beyond the water lines that dominate their spectra, while the spectra of the warmest M dwarfs have a large number of measurable Fe I lines and very few measurable FeH lines.
On average, we find an offset between the abundances calculated from Fe I and FeH, with the Fe I lines giving Fe abundance values that are systematically higher than FeH lines by roughly 0.05 dex. For the seven M dwarfs in our sample we obtain: $\langle$A(Fe I)$\rangle$ = 7.51 $\pm$ 0.04 dex, and $\langle$A(FeH)$\rangle$ = 7.47 $\pm$ 0.06 dex.
The offset between the iron abundances from Fe I and FeH are small and, in part, could be the result of systematic uncertainties in the log $g$f values of FeH lines since these were computed from intensities, while the $g$f values for the Fe I lines were adjusted to match the abundances of the benchmark stars Sun and Arcturus. See \cite{Smith2021} for details on the construction of the APOGEE line list.

\subsection{Comparisons with APOGEE DR16 and Literature Results}

\begin{figure}
	\includegraphics[width=1\linewidth]{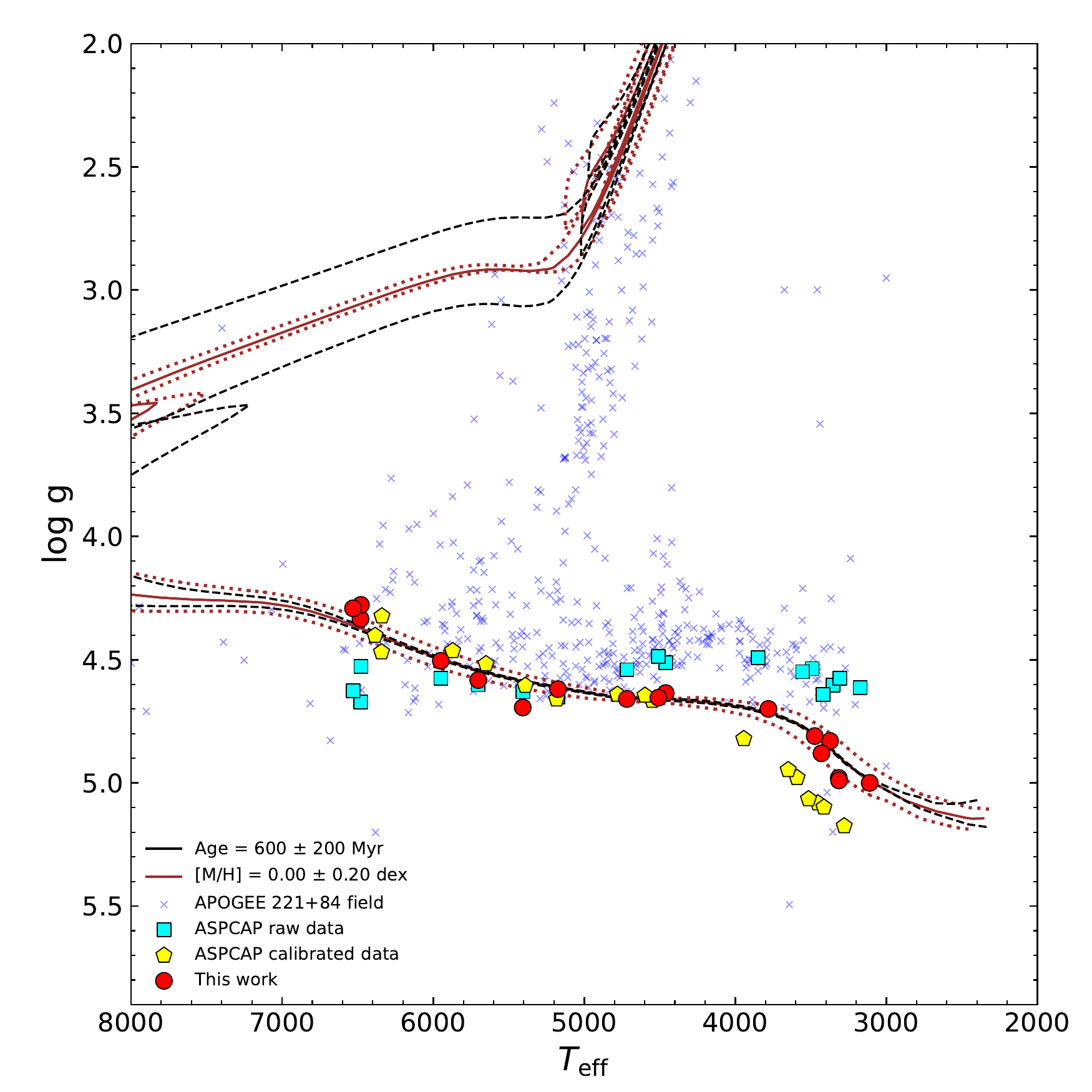}
    \caption{The Kiel ($T_{\rm eff}$--log $g$) diagram for this work studied sample shown as red circles and the APOGEE targets in the 221+84 field are shown as blue crosses. MIST isochrones are shown as brown and black lines.}
    \label{Kiel}
\end{figure}

In Figure \ref{Kiel} we show a Kiel diagram with the same MIST isochrones for an age = 600 $\pm$ 200 Myr and metallicity [Fe/H] = 0.00 $\pm$ 0.20 as in Figure \ref{membership1}; the stellar parameters adopted for the target stars (discussed in Section 3.1) are shown as red circles and these show good consistency with respect to the MIST isochrones. 
For comparison, we also show the $T_{\rm eff}$ and, in particular, log $g$ values from DR16: both calibrated (yellow pentagons) and raw uncalibrated (cyan squares). 
The ASPCAP uncalibrated log $g$s are roughly constant, falling close to the MIST isochrones only in the limited interval roughly between $T_{\rm eff}$ = 5000 K and 6000 K, outside of this interval, there is a clear disagreement in the log $g$ values for stars cooler and hotter than this $T_{\rm eff}$ range; for $T_{\rm eff}$ $>$ 6000 K, the uncalibrated ASPCAP parameters fall below, while for lower $T_{\rm eff}s$ these are systematically above the MIST isochrones.

To try to mitigate this surface gravity offset, the APOGEE team applies calibrations to the ASPCAP log $g$, which are based on the more precise log $g$s obtained from asteroseismology for stars in the Kepler field \citep{Pinsonneault2018,Serenelli2017}.
However, for the M dwarf regime, asteroseismic log $g$s are not available, and the log $g$ calibration relies instead on the Parsec isochrones \citep{Bressan2012}.
The ASPCAP effective temperatures are also calibrated, and the APOGEE team uses the photometric $T_{\rm eff}$-scale from \cite{GonzalezHernandez2009} for a sample of stars with no, or very low, reddening \citep{Holtzman2018}. 
It is worth pointing out, however, that the uncalibrated ASPCAP $T_{\rm eff}$-scale for F, G, K dwarfs compares well with the effective temperatures obtained by \cite{Cintia2019} for a sample of F, G, K dwarf hosting planets from the California Kepler Survey \citep{Fulton2018}. Their results are from a spectroscopic analysis of optical Fe I and Fe II lines from Keck HIRES spectra. 
The two completely independent spectroscopic $T_{\rm eff}$-scales are in good agreement: $\delta$ $T_{\rm eff}$ ($\langle$ ASPCAP raw - \cite{Cintia2019} $\rangle$) = -68.7 $\pm$ 124.0 K (rms) for a sample of 490 stars in common between APOGEE DR16 and \cite{Cintia2019}.
Overall, Figure \ref{membership1} shows that the ASPCAP DR16 calibrated stellar parameters (yellow pentagons) are much more consistent with the MIST isochrones, except for the M dwarf regime ($T_{\rm effs}$ $<$ 4000 K). However, much of this improvement comes from the calibrations in surface gravities, which are much more significant.

\begin{figure*}
	\includegraphics[width=.33\linewidth]{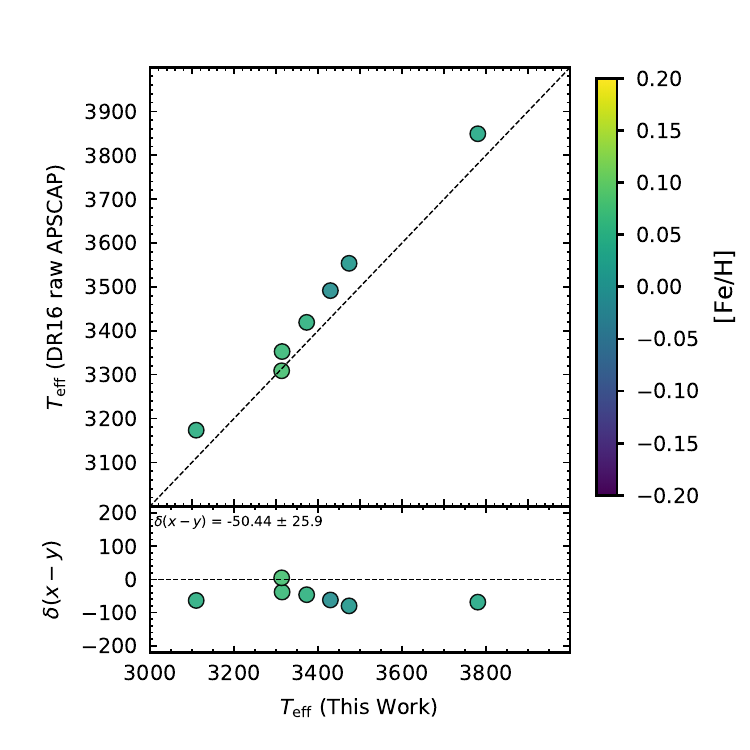}
	\includegraphics[width=.33\linewidth]{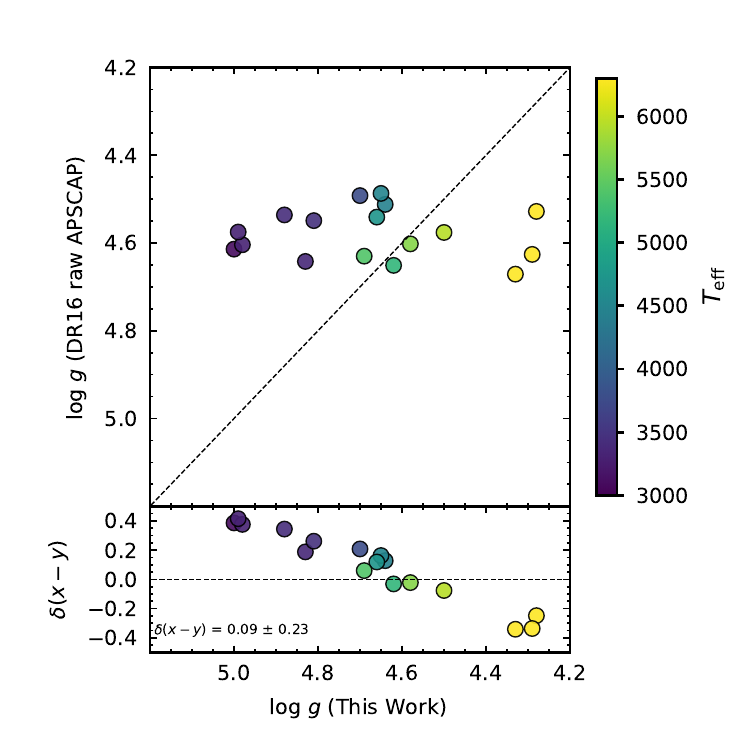}
	\includegraphics[width=.33\linewidth]{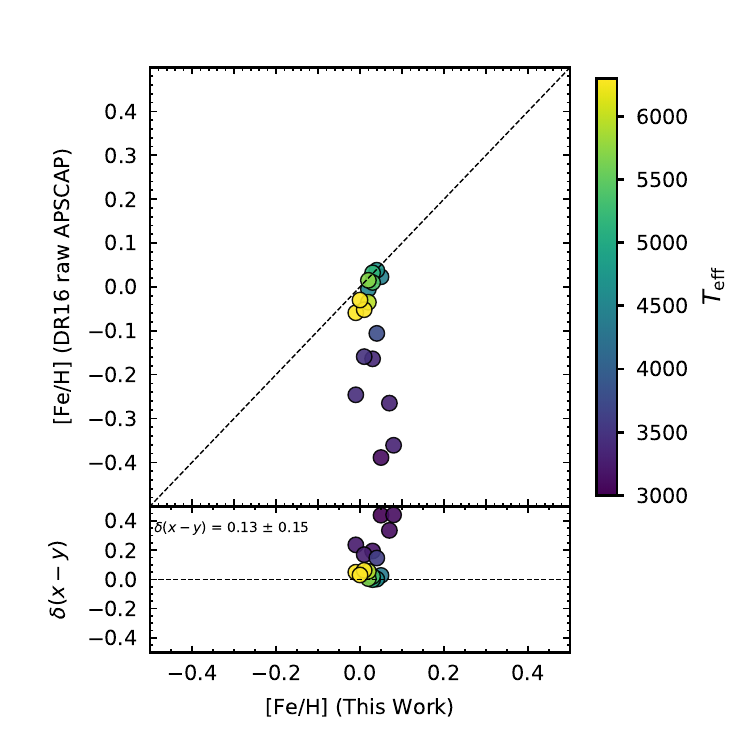}
	\includegraphics[width=.33\linewidth]{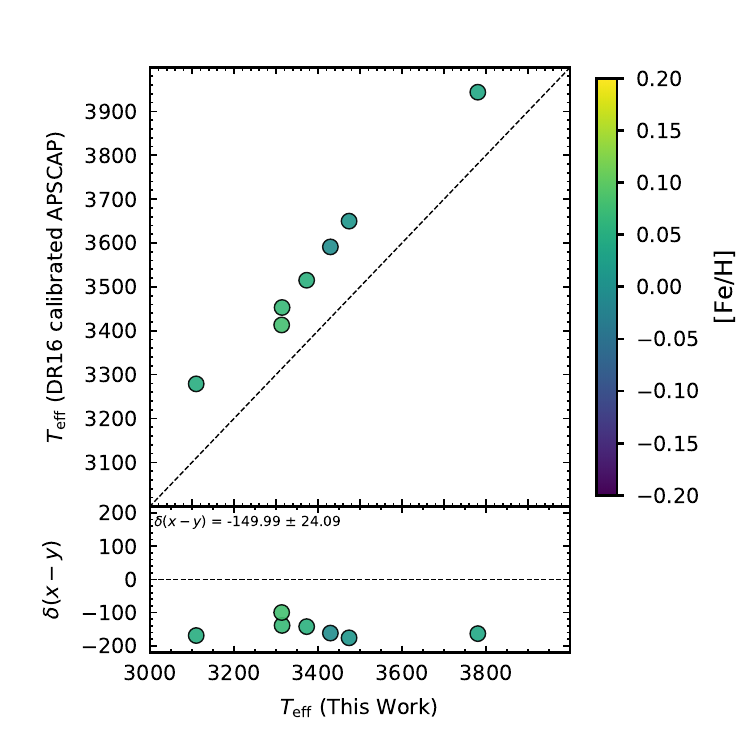}
    \includegraphics[width=.33\linewidth]{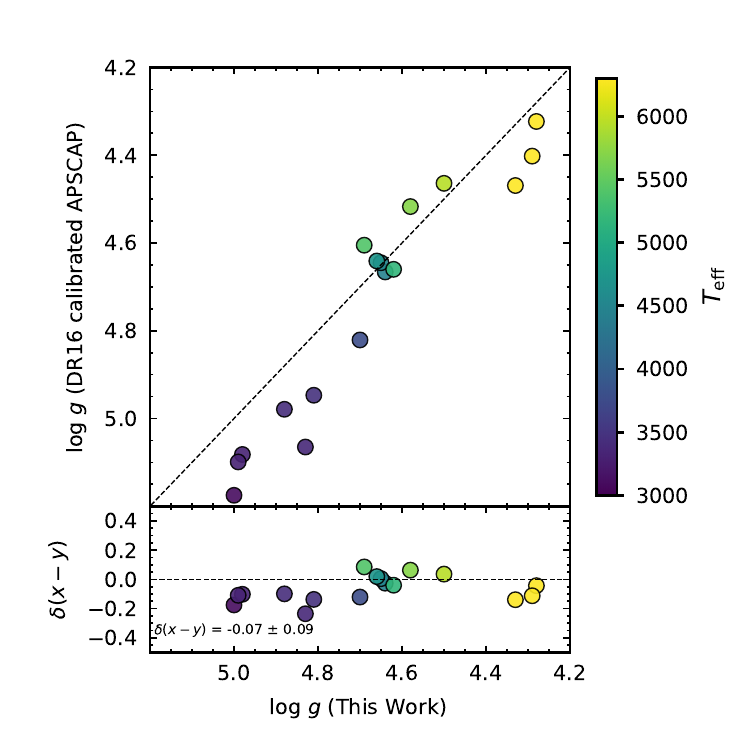}
	\includegraphics[width=.33\linewidth]{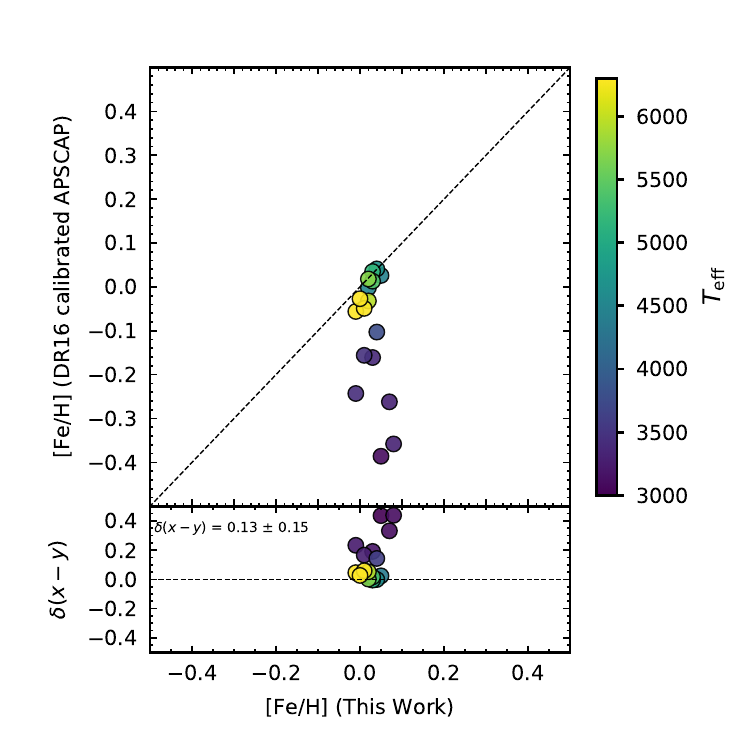}
    \caption{The $T_{\rm eff}$ (left panel), log $g$ (middle panel) and, [Fe/H] (right panel) diagrams comparing this work with the ASPCAP DR16 raw (upper panels) and calibrated (bottom panels) results. A residual diagram is shown in the bottom of each panel.}
    \label{DR16}
\end{figure*}

A comparison of the results in this study with the ASPCAP DR16 $T_{\rm eff}$ and log $g$ results, both calibrated and uncalibrated (raw) are presented in the different panels of Figure \ref{DR16}; residual diagrams are shown at the bottom of each panel.
The top left panel of Figure \ref{DR16} shows that there is good agreement between the effective temperature scales for the M dwarfs derived here with those from the ASPCAP DR16 results, where $T_{\rm eff}$ (This work - ASPCAP raw) = -50 $\pm$ 26 K. Such difference and rms are very small, but it is puzzling that the calibrated effective temperatures for the M dwarfs have a much poorer agreement with our results; the ASPCAP calibrated effective temperatures for M dwarfs (shown in the bottom left panel of Figure \ref{DR16}) are systematically higher than ours ($T_{\rm eff}$ (This work - ASPCAP calibrated) $\sim$ -150 K $\pm$ 24 K). Taken at face value, our results may hint that the effective temperatures of M dwarfs were not adequately calibrated in DR16. 
For the log $g$ (middle panels of Figure \ref{DR16}), the ASPCAP raw results are clumped around log $g$ = 4.6 dex, while the physical log $g$s computed here, which were derived from a fundamental relation (equation 1), varies between $\sim$ 4.0 to 5.0 dex. There is a clear trend in the log $g$ difference: the ASPCAP raw log $g$s are systematically lower than ours for the cooler stars and systematically higher than ours for the warmer stars.
The calibrated ASPCAP log $g$s show offsets that on average are not too large: $\delta$log $g$ (this work - ASPCAP calibrated) = -0.14 $\pm$ 0.05 dex, but the differences change sign at log $g$ $\sim$ 4.6.

\begin{deluxetable*}{llccccccc}
\tablecaption{M dwarfs line-by-line Abundances\label{M_abu}}
\tablewidth{700pt}
\tabletypesize{\scriptsize}
\tablehead{
Element	& Lambda 	&J12193796+	&J12264027+	&J12201448+	&J12231356+	&J12255421+	&J12250262+	&J12241121+\\
	&  (\AA{})	&2634445	&2718434	&2526072	&2602185	&2651387	&2642382	&2653166}
\startdata
FeI&	15194.5&	...&	...&	...&	...&	...&	...&	...\\
&	15207.5&	...&	...&	7.44&	...&	...&	7.48&	7.58\\
&	15219.5&	...&	...&	...&	...&	...&	...&	...\\
&	15244.8&	...&	...&	...&	...&	...&	...&	7.60\\
&	15395.0&	...&	...&	...&	...&	...&	...&	...\\
&	15490.3&	...&	7.48&	...&	...&	7.35&	7.42&	7.48\\
&	15591.8&	...&	...&	...&	...&	...&	...&	7.53\\
&	15604.0&	...&	...&	...&	...&	...&	...&	7.61\\
&	15621.7&	...&	...&	7.46&	&	7.58&	7.60&	7.64\\
&	15632.0&	...&	7.56&	7.51&	7.52&	7.48&	7.53&	7.61\\
&	15648.5&	...&	...&	...&	...&	...&	...&	7.56\\
&	15662.0&	...&	...&	...&	...&	...&	...&	7.60\\
&	15692.5&	...&	...&	...&	...&	...&	...&	7.68\\
&	15723.5&	...&	...&	...&	...&	...&	...&	7.58\\
&	&	&	&	&	&	&	&	\\
$\langle$A(Fe)$\rangle$    &   &	...&	7.52&	7.47&	7.52&	7.47&	7.51&	7.59\\
$\langle$[FeI/H]$\rangle$   &   &	...&	0.07&	0.02&	0.07&	0.02&	0.06&	0.14\\
$\sigma$ &   &	...&	0.06&	0.04&	...&	0.12&	0.08&	0.05\\
\hline
FeH&	15965.0&	...&	7.60&	7.59&	7.68&	7.42&	7.47&	7.44\\
&	16009.6&	...&	...&	...&	...&	...&	...&	7.56\\
&	16018.5&	7.50&	...&	7.43&	7.59&	7.15&	7.13&	...\\
&	16108.1&	...&	...&	7.61&	...&	7.56&	7.51&	7.30\\
&	16114.0&	...&	7.54&	7.42&	7.55&	7.40&	7.37&	7.33\\
&	16245.7&	...&	7.62&	7.42&	7.66&	7.29&	7.36&	7.32\\
&	16271.8&	...&	7.63&	7.58&	7.61&	7.46&	7.48&	7.32\\
&	16284.7&	7.50&	7.31&	7.46&	7.34&	7.33&	7.35&	7.46\\
&	16299.4&	...&	...&	...&	...&	...&	...&	...\\
&	16377.4&	...&	...&	7.50&	7.46&	7.40&	7.43&	7.33\\
&	16546.8&	...&	7.49&	7.48&	7.52&	7.42&	7.44&	7.34\\
&	16548.8&	...&	7.44&	7.37&	7.51&	7.40&	7.38&	\\
&	16557.2&	&	...&	7.35&	7.31&	7.31&	7.38&	7.40\\
&	16574.8&	...&	7.58&	7.56&	7.69&	7.48&	7.45&	7.42\\
&	16694.4&	...&	7.59&	7.51&	7.55&	7.51&	7.46&	7.43\\
&	16735.4&	...&	...&	...&	...&	...&	...&	...\\
&	16738.3&	...&	7.45&	...&	7.55&	...&	...&	...\\
&	16741.7&	...&	7.56&	7.49&	7.59&	...&	...&	...\\
&	16796.4&	...&	7.55&	7.56&	7.58&	7.45&	7.45&	...\\
&	16812.7&	...&	7.56&	7.51&	7.57&	7.46&	7.41&	...\\
&	16814.1&	...&	7.53&	7.47&	7.56&	7.51&	7.46&	...\\
&	16889.6&	...&	7.46&	7.59&	7.57&	7.51&	...&	...\\
&	16892.9&	...&	7.32&	7.44&	7.52&	...&	...&	...\\
&	16922.7&	...&	...&	...&	7.35&	...&	...&	...\\
&	16935.1&	...&	7.61&	7.55&	7.49&	...&	...&	...\\
&	&	&	&	&	&	&	&	\\
$\langle$A(FeH)$\rangle$&	&	7.50&	7.52&	7.49&	7.54&	7.42&	7.41&	7.39\\
$\langle$[FeH/H]$\rangle$&	&	0.05&	0.07&	0.04&	0.09&	-0.03&	-0.04&	-0.06\\
$\sigma$&	&	0.00&	0.10&	0.08&	0.10&	0.10&	0.09&	0.08\\
\enddata
\end{deluxetable*}

In the right panels of Figure \ref{DR16}, we show the comparison with the ASPCAP metallicities. The first point is that the metallicities from ASPCAP, both calibrated and uncalibrated, are quite similar. However, we can see that there are some significant differences with our results.
To simply compute an average difference for the full studied sample is not very meaningful because the differences in the metallicity results are very $T_{\rm eff}$ dependent. The coolest M dwarfs (represented by the darker blue circles according to the color bar in the figure) have metallicities that are lower by roughly 0.4 dex when compared with our results. This is a significant difference that points to still unresolved issues in the ASPCAP results for the cool M dwarfs. \citep[See also ][ who also find a discrepancy in metallicity compared to ASPCAP]{Sarmento2021}.
There is much better agreement with the metallicities obtained by ASPCAP for the warmer FGK stars, $\delta$[Fe/H] (this work - ASPCAP raw) = 0.03 $\pm$ 0.02 dex, such difference is within the expected uncertainties given the different analysis methods and the metallicities of the G stars in our study should represent a solid anchor for the abundances of the Coma Ber open cluster.

The metallicity discrepancy of the ASPCAP DR16 results for the M dwarfs is likely to be driven by the selection of the Fe I lines used to define the spectral windows for the abundance determinations that were constructed based on the spectra of the benchmarks Arcturus and the Sun. 
APOGEE adopts the same spectral windows to derive chemical abundances for the entire APOGEE survey that has observed well over 650,000 stars, of which just a small minority are M dwarfs.
The APOGEE spectrum of an M dwarf is significantly different from that of a red giant or solar-type star due to the presence of FeH lines and H$_{2}$O that dominate the spectra at cool temperatures. At this time, the APOGEE team is working on setting specific analysis methods to analyze this stellar class properly in the future and mitigate this problem in the final APOGEE data release.

Our derived metallicities can also be compared with results from optical studies from the literature.
One of the first high-resolution spectroscopic studies of stars in Coma Ber was by \cite{FrielBoesgaard1992}. Their analyzed sample was composed of FG dwarfs, and they obtained an average metallicity of [Fe/H] = -0.052 dex for this open cluster, which is just slightly metal-poor relative to the Sun. Some of the other abundance studies of Coma Ber stars conducted spectroscopic analyses of AF type stars \citep{Hui-Bon-Hoa1997,Hui-Bon-Hoa1998,Gebran2008}, while other works determined the cluster metallicity using photometry and isochrones \citep{Paunzen2010,Netopil2016}; all of these confirming the metallicity scale of Coma Ber to be roughly solar.

\begin{figure}
	\includegraphics[width=1\linewidth]{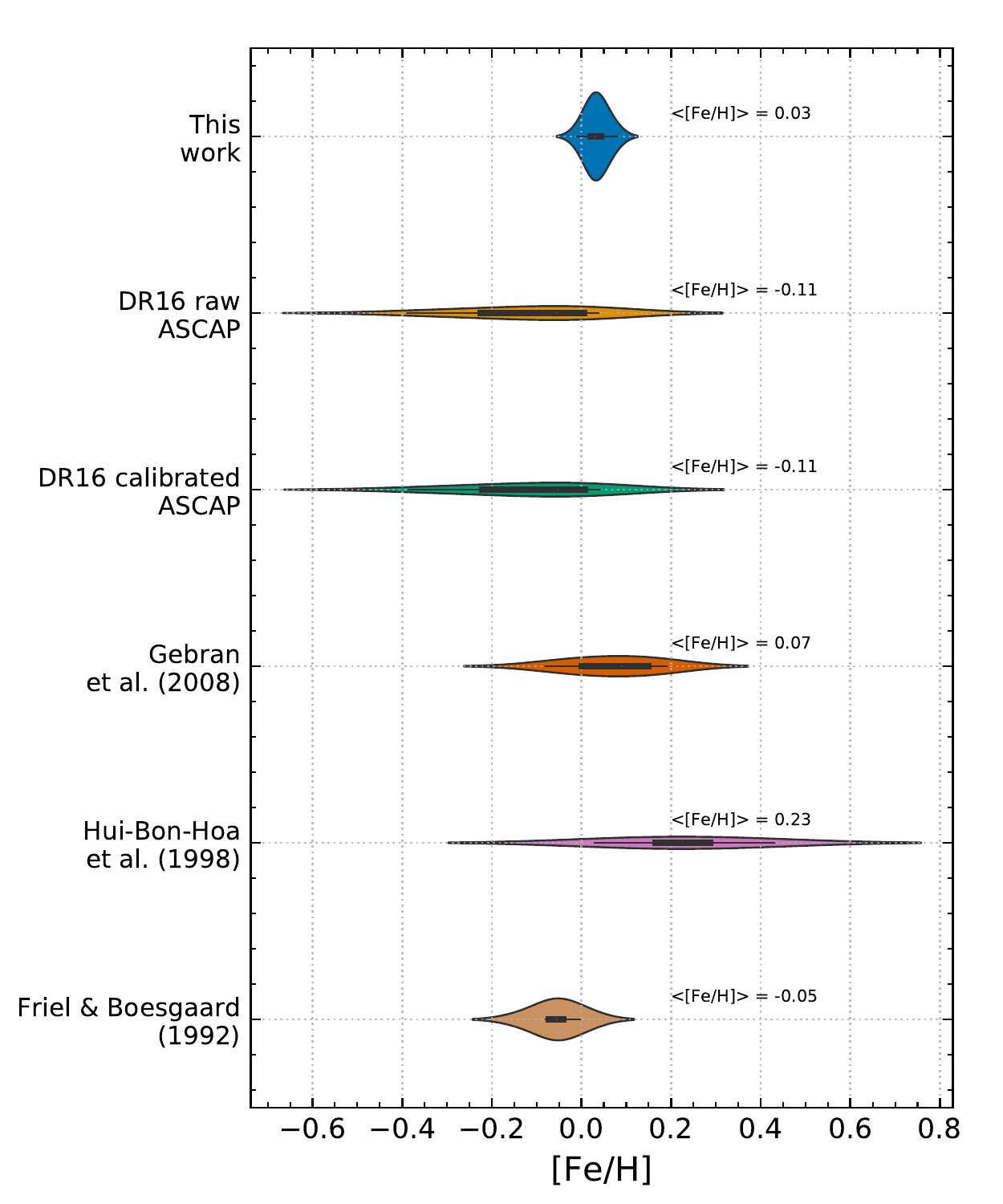}
    \caption{Metallicity distribution of Coma Ber stars of this work compared with the literature. }
    \label{lit_comp}
\end{figure}

In Figure \ref{lit_comp} we show violin diagrams for the Coma Ber metallicities from different spectroscopic studies. From top to bottom, we display the metallicity diagrams from this work, the raw APOGEE DR16 results, the calibrated APOGEE results (which are almost identical to the raw ones), and results from \cite{Gebran2008}, \cite{Hui-Bon-Hoa1998}, and \cite{FrielBoesgaard1992}. 
In general, the mean metallicity obtained here is in reasonable agreement with that from \cite{Gebran2008} ($\delta$[Fe/H] (This work - \cite{Gebran2008}) = -0.03 dex) and our results are also marginally consistent with those from \cite{FrielBoesgaard1992} ($\delta$[Fe/H] (This work - \cite{FrielBoesgaard1992}) = 0.09). 

The mean metallicity for Coma Ber obtained by \cite{Hui-Bon-Hoa1998} ($<$[Fe/H]$>$=+0.23) is significantly more metal-rich than ours, and the distribution in that study also shows a much larger spread.
The APOGEE DR16 results, on the other hand, are overall more metal-poor ($<$[Fe/H]$>$ = -0.11) and also show a much larger spread when compared to ours.
We should note that the violin diagram for the APOGEE DR16 results and those from our study correspond to the same stars, analyzing the same APOGEE spectra, using the same LTE radiative transfer code and MARCS model atmospheres grid, and adopting the same APOGEE line list, but using different methodologies and diagnostic lines to derive the metallicities. As discussed above, this discrepancy in the metallicity distribution is mostly caused by differences in the metallicity results for the cooler stars. 
We note that we exclude the F star results in this metallicity comparison due to the presence of atomic diffusion processes, as discussed in the following Section.

\subsection{Chemical Homogeneity and Diffusion in Coma Ber Stars}

Quantifying any chemical abundance differences between member stars in a cluster is an important exercise in both setting constraints on the chemical homogeneity within the molecular cloud that formed the cluster, which limits the ability of chemical tagging \citep{Freeman2002} to connect particular stars to particular clusters via abundance patterns, as well as surface abundance variations related to stellar evolution. 
There is a wealth of previous chemical abundance studies in the literature that investigated chemical homogeneity in open clusters via optical spectroscopy \citep[e.g., ][]{Michaud2004,DeSilva2006,DeSilva2007,Carrera2013,Feng2014,Blanco-Cuaresma2015,Liu2016b,BertelliMotta2018,Gao2018,Casamiquela2020}.
The APOGEE OCCAM survey observed a large sample of open clusters \citep{Donor2020} and APOGEE results were also used to test chemical homogeneity in clusters \citep{Bovy2016,BertrandeLis2016,GarciaDias2019,Poovelil2020,Cheng2020}. 
Most of these previous studies found that the open clusters had homogeneous chemical abundances to a 2- to 3-sigma confidence level, at a few-hundreths-of-a-dex, when considering stars from a single, narrow evolutionary state.  Comparisons within cluster member stars that span a range of evolutionary stages, such as main sequence stars compared to red giants, have measured abundance variations caused by stellar evolution during red giant first dredge-up, which results in an increase in $^{14}$N abundances, along with small decreases in $^{12}$C abundances, as well as dramatically lower $^{12}$C/$^{13}$C ratios along the red giant branch (RGB) \citep[e.g., ][]{Szigeti2018}.

An additional process that may modify surface abundances in main sequence stars is atomic diffusion, which can transport material in the stellar atmosphere via diffusive processes, such as gravitational settling or radiative acceleration. The magnitude of abundance variations created by atomic diffusion is mainly a function of stellar metallicity, mass, and age. Also, atomic diffusion processes operate more efficiently in radiative regions \citep{Michaud2004,Dotter2017}, therefore, more easily detectable in certain temperature regimes on the main sequence in open clusters.
In general, for old metal-poor stars ([Fe/H] $\sim$ -2.00 dex; Age $\sim$ 10 Gyr) there has been sufficient time for gravitational settling to occur, as observed, for example, in the metal-poor globular cluster NGC 6397 by \cite{Korn2007}, while for young metal-rich stars ([Fe/H] $\sim$ 0.00 dex; Age $<$ 1 Gyr) gravitational settling is still in its earlier stages \citep{Michaud2005}, resulting in smaller changes in abundance.

Using APOGEE spectra, \cite{Souto2018,Souto2019a} found chemical inhomogeneities in the stars belonging to the M67 open cluster ([Fe/H] $\sim$ 0.00; Age $\sim$ 4.0 Gyr).
They found that the turnoff stars in M67 had systematically lower metallicities, by roughly 0.15 dex, when compared to stars in other evolutionary stages, and that stellar models which included atomic diffusion described well the observed abundance variations. 
Stellar abundance variations across the Hertzsprung–Russell (HR) diagram of M67 were also observed using optical spectra from the Gaia-ESO survey \citep{BertelliMotta2018}, and from the GALAH survey \citep{Gao2018}, and in \cite{Onehag2014}. 
Recently, \cite{Semenova2020} found the signature of atomic diffusion in the open cluster NGC 2420, which is about half of the M67 age \citep[$\sim$ 2Gyr]{Sharma2006} and has slightly sub-solar metallicity according to the APOGEE study by \citep{Souto2016}.

The Coma Ber stars in which iron abundances have been derived are younger than stars in both M67 and NGC 2420 and can be used to probe diffusion in main sequence stars with ages near 600 Myr.  The effective temperature range of the Coma Ber cluster stars includes $T_{\rm eff}$ $\sim$ 6500 K, at the hot end, where the convective zone mass is quite small, M$_{\rm CZ}\sim$0.003M$_{\odot}$ \citep{Choi2016}, and abundance changes due to diffusion might be measurable.  Diffusion effects likely weaken in the progressively cooler G and K dwarfs, and would be expected to vanish in the cool, convective M dwarfs.  As the APOGEE sample contains F, G, K, and M main sequence stars, it is a useful sample in which to test chemical homogeneity and diffusion.

\begin{figure}
	\includegraphics[width=0.9\linewidth]{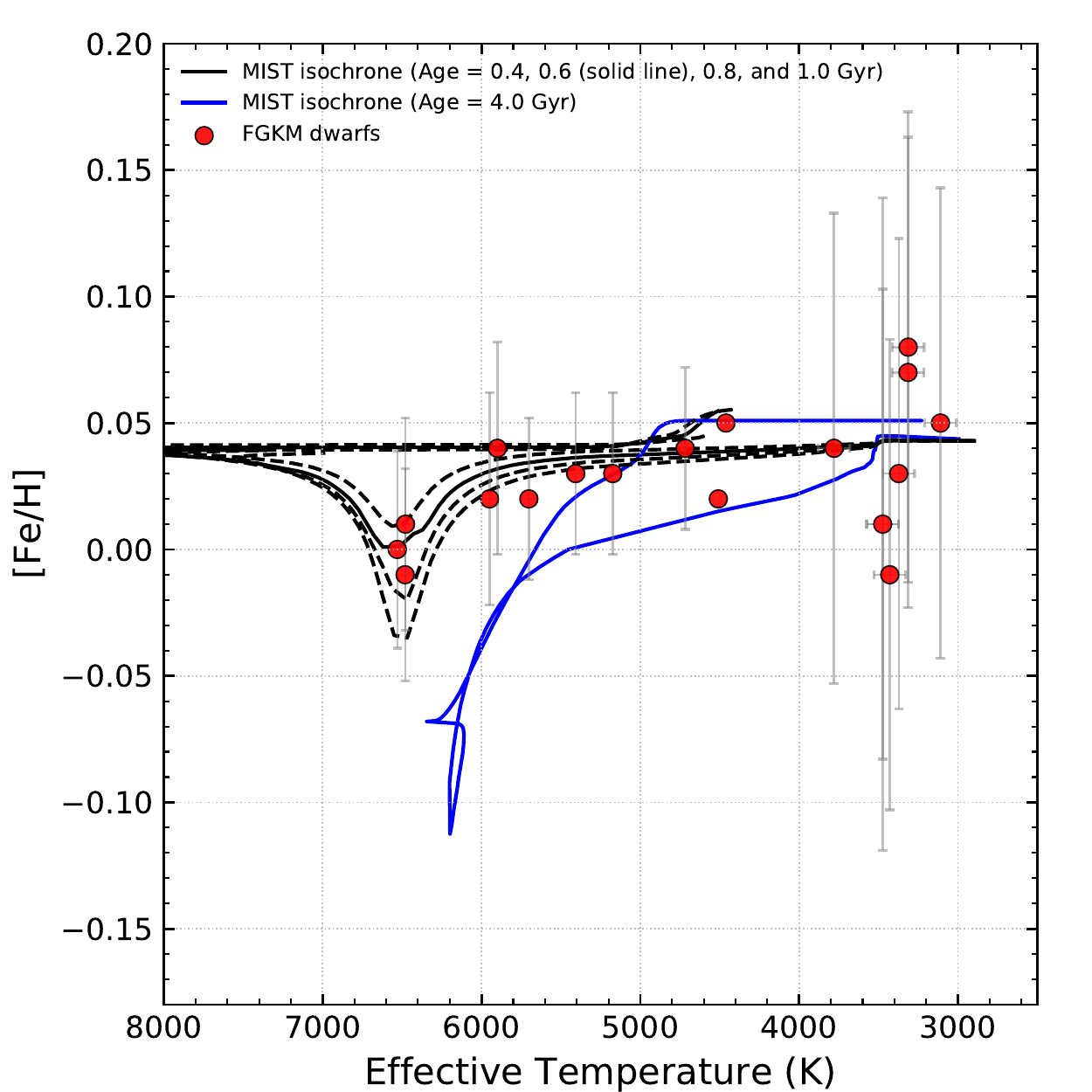}
	\includegraphics[width=0.9\linewidth]{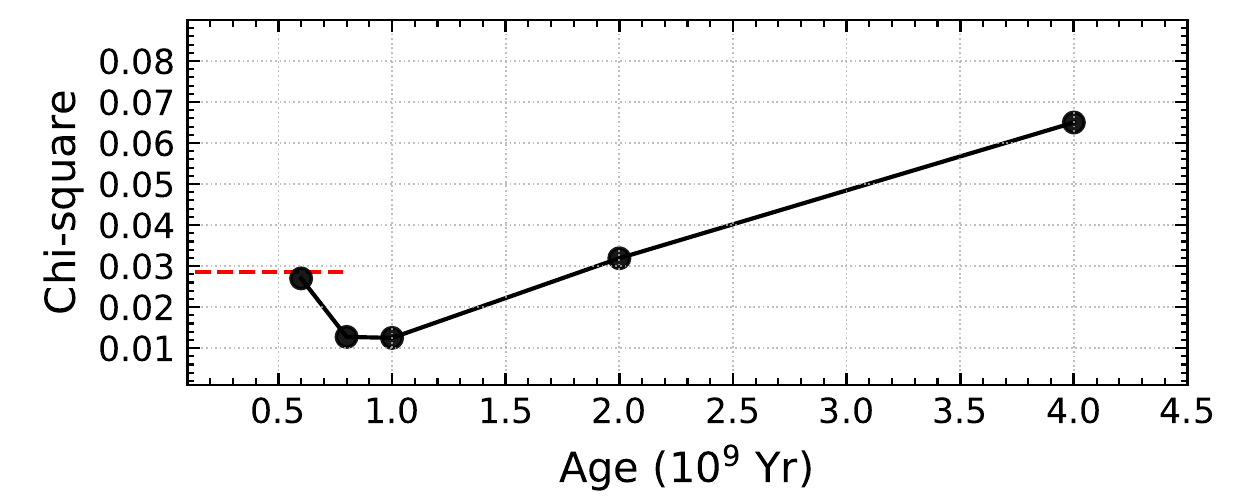}
    \caption{Top panel: the F, G, K, M dwarf metallicities as a function of $T_{\rm eff}$. We show the adopted MIST isochrone assuming [Fe/H] = 0.00 dex and age = 0.4, 0.6, 0.8, and 1.0 Gyr as black solid and dashed lines, and as a solid blue line, we show the 4.0 Gyr isochrone for the same solar metallicity.} The bottom panel shows the chi-square minimization comparing the derived metallicities with MIST isochrones. The red dashed line indicates the chi-square compared to the mean metallicity derived from Coma Ber.
    \label{teff_met}
\end{figure}

The top panel of Figure \ref{teff_met} presents [Fe/H] as a function of $T_{\rm eff}$ for the Coma Ber stars plotted as filled red circles (the error bars represent the uncertainties as described in Table \ref{uncert}).  The continuous black curves shown in Figure \ref{teff_met} (top panel) are three atomic diffusion models from MIST isochrones \citep{Choi2016}, assuming [Fe/H] = 0.00 and ages of 600 Myr (solid curve), plus 400 Myr and 800 Myr (dashed curves).  The horizontal line represents the mean Coma Ber [Fe/H] = 0.04 as defined by the G, K, and M dwarfs, where the convective zone masses are increasing; the diffusion models are also shifted to this value of [Fe/H] as a starting point.  Also included in the top panel of Figure \ref{teff_met} is the diffusion model for an age of 4.0 Gyr (typical of M67), which has quite a different structure than the younger Coma Ber-like diffusion isochrones.  Of note is the deeper diffusion signature in [Fe/H] along the main sequence and, in particular, the deep dip just before the M67 turnoff, near $T_{\rm eff}$ = 6400 K (or turnoff mass, M$\sim$1.2M$_{\odot}$).  The iron abundances in the 4.0 Gyr model then increase in stars that have evolved past the turnoff as the deepening convective envelopes of the sub-giant and giant stars erase the main-sequence/turnoff diffusion signature.  The behavior of the [Fe/H] dip in the younger Coma Ber-like models spans a relatively restricted range in $T_{\rm eff}$ $\sim$ 6200-6800 K (half-depth), with the decreasing diffusion signature (and eventual disappearance) at the higher effective temperatures caused by the increasing effectiveness of radiative processes.  Due to its younger age, any turnoff stars in Coma Ber are hotter than shown in Figure \ref{teff_met} (top panel) and not analyzed here: they within the spectral-type range of A stars.  The derived values of [Fe/H] in the Coma Ber stars exhibit scatter, with the individual uncertainties of the F, G, and K dwarfs ($T_{\rm eff}$ $>$ 4000 K) being comparable to the magnitude of the diffusion signature ($\sim$0.04 dex).  Even given the individual uncertainties in [Fe/H], the trend over the range of $T_{\rm eff}$ = 4000 K to 6500 K follows the diffusion models covering the approximate age of Coma Ber.  The M dwarfs exhibit the largest abundance uncertainties (Table \ref{uncert}) and scatter, although their mean abundances agree well with an underlying pristine abundance of [Fe/H] = +0.04.

A quantitative comparison between the derived stellar values of [Fe/H] with the diffusion models (all shown in the top panel of Figure \ref{teff_met}) is presented in the bottom panel of Figure \ref{teff_met} as a $\chi^{2}$ difference as a function of model age.  Diffusion models for ages of 600 Myr, 800 Myr, 1.0 Gyr, 2.0 Gyr, and 4.0 Gyr were used in the comparison fits.  The dashed horizontal red line shows the $\chi^{2}$ value for a constant mean abundance of [Fe/H] = +0.04.  the diffusion models with ages $\sim$800 Myr - 1 Gyr yield the best fits and provide some evidence that the signature of diffusion has been detected in the [Fe/H] abundances along the main sequence of Coma Ber.

Coma Ber members have also been analyzed for Li and Be by \cite{Boesgaard1987} and \cite{Boesgaard2003}, respectively, in main-sequence stars from $T_{\rm eff}$ $\sim$ 5700 K - 8500 K.  \cite{Boesgaard1987} Li abundances span the so-called ``Li dip'', which becomes detectable at $T_{\rm eff}$ $\sim$ 6400 K, reaches a maximum depth at $T_{\rm eff}$ $\sim$ 6600 K, and then returns to the cluster's pristine Li abundance by about $T_{\rm eff}$ $\sim$ 6900 K; this study includes four A-type main-sequence stars ($T_{\rm eff}$ = 7900 K-8500 K), which display a mean lithium abundance of A(Li) = 3.1 dex, close to the pristine Li abundance in Coma Ber. 
The Li dip defined by \cite{Boesgaard1987} analysis coincides approximately with the small dip in the metallicities observed here and the models by \cite{Choi2016}. The beryllium abundances from \cite{Boesgaard2003} also exhibit a dip in the Be abundances, which also coincides, approximately, with the Li dip. 

The Li dip has been studied in several open clusters (e.g., see \citealt{Cummings2012} and references therein) with, in particular, \cite{Cummings2017} presenting a detailed analysis and mapping of the Li dip in the $\sim$650 Myr old, metal-rich open clusters the Hyades and Praesepe. \cite{Cummings2017} investigate both diffusion models, as well as rotationally-induced mixing models as possible causes of the Li dip and discuss evidence that indicates that, perhaps, both processes might be responsible for sculpting the shape of the Li dip.  This suggestion impacts the interpretation of the Fe abundances presented here, and it is worth noting that rotationally-induced mixing is also included in the models by \cite{Choi2016} as a type of ``diffusion'', along with gravitational and radiative accelerations.  The addition of abundances from a wider range of elements, which have different diffusion coefficients, might provide new insights into the relative amounts of abundance variations that could be produced by gravitational and radiative diffusion when compared to rotationally-driven mixing.
 
\section{Conclusions}

The APOGEE survey observed a number of Galactic open clusters \cite{Donor2020} and systematically targeted stars in the field of the young \citep[600 Myr old, ][]{GaiaCollaboration2018_HRdiag} and nearby \cite[d $\sim$ 80 pc, ][]{Tang2018comaber} open cluster Coma Berenices. 
Using the membership study of \cite{Tang2018comaber}, Gaia DR2 \citep{GaiaCollaborationDR2} and APOGEE radial velocities \citep{Majewski2017}, in this work, we selected a sample of seven M-, four K-, four G- and three F-dwarf bona fide members of Coma Ber and determined their metallicities using the high-resolution, high-signal-to-noise NIR spectra of the APOGEE survey. 
This is the first detailed metallicity study including APOGEE M dwarf stars belonging to an open cluster. The studied Coma Ber M dwarfs cover an extended range in effective temperature from $\sim$ 3100 - 3900 K, offering the opportunity to probe systematic differences in the results as a function of $T_{\rm eff}$. Following the methodology in previous studies of APOGEE M dwarfs \citep{Souto2020} we use an independent analysis method when compared to that adopted, for example, in the APOGEE automatic abundance pipeline ASPCAP; the stellar parameters and metallicities derived here are based upon measurements of  H$_{2}$O and OH lines and Fe I and FeH lines, while the ASPCAP pipeline fits the entire APOGEE spectrum and uses only Fe I spectral windows to derive metallicities.

In overall agreement with what was previously found in the literature, our results for the sample of G, K, and M dwarfs indicated that Coma Ber has near-solar metallicity; the mean iron abundance and standard deviation of the mean is $\langle$[Fe/H]$\rangle$ = 0.04 $\pm$ 0.02 dex. Moreover, when breaking the sample into bins in effective temperature (roughly corresponding to spectral types) the metallicities obtained are found to be quite homogeneous, suggesting that this open cluster can be well characterized by a single Fe abundance, a result that is in line with the paradigm of chemical tagging \citep{Freeman2002}. 
We find: 
$\langle$[Fe/H]$\rangle$ = 0.04 $\pm$ 0.03 dex for the M dwarfs, 
$\langle$[Fe/H]$\rangle$ = 0.04 $\pm$ 0.01 dex for the K dwarfs, 
$\langle$[Fe/H]$\rangle$ = 0.03 $\pm$ 0.01 dex for the G dwarfs, and $\langle$[Fe/H]$\rangle$ = 0.00 $\pm$ 0.01 dex for the F dwarfs.

Although the differences between mean values of [Fe/H] for the M, K, G, and F dwarfs noted above are small, detailed comparisons of the derived iron abundances with models incorporating atomic diffusion, as a function of $T_{\rm eff}$, reveal the operation of chemical diffusion over a timescale close to the adopted age of Coma Ber. The magnitude of the abundance decrease of $\sim$0.04 dex near $T_{\rm eff}$ = 6500 K is in general agreement with the models of \cite{Choi2016} for MIST isochrones corresponding to ages near $\sim$800 Myr. The comparisons of the Fe abundance dip observed here with the results on the Li and Be dips observed by \cite{Boesgaard1987} and \cite{Boesgaard2003}, respectively, may suggest that the effects of atomic diffusion have been observed in the stars of Coma Ber, specifically in the F stars in this study, while the metallicities of the cooler, lower-mass K and M dwarfs should represent the pristine metallicity of the Coma Ber open cluster. In addition, it is possible that rotational-induced mixing has also affected the metallicity dip observed here (e.g., \citealt{Cummings2017}).

\section*{Acknowledgements}

We wholehearted thank the APOGEE team and the APOGEE PI Steve Majewski for all the work done to accomplish the APOGEE survey. This paper used publicly available APOGEE data. 
We thank Marc Pinsonneault for discussions about diffusion. 
We thank the anonymous referee for useful comments that helped improve the paper. 
We thank Ryan Terrien and Suvrath Mahadevan for helping in the planning of the APOGEE observations of Coma Ber.
KC acknowledges partial support by the National Aeronautics and Space Administration under Grant 18-2ADAP18-0113. KC and VS acknowledge that their work here is supported, in part, by the National Aeronautics and Space Administration under Grant 16-XRP16\_2-0004, issued through the Astrophysics Division of the Science Mission Directorate. 

Funding for the Sloan Digital Sky Survey IV has been provided by the Alfred P. Sloan Foundation, the U.S. Department of Energy Office of Science, and the Participating Institutions. SDSS-IV acknowledges
support and resources from the Center for High-Performance Computing at the University of Utah. The SDSS web site is www.sdss.org.

SDSS-IV is managed by the Astrophysical Research consortium for the 
Participating Institutions of the SDSS Collaboration including the 
Brazilian Participation Group, the Carnegie Institution for Science, 
Carnegie Mellon University, the Chilean Participation Group, the French Participation Group, Harvard-Smithsonian Center for Astrophysics, 
Instituto de Astrof\'isica de Canarias, The Johns Hopkins University, 
Kavli Institute for the Physics and Mathematics of the Universe (IPMU) /  
University of Tokyo, Lawrence Berkeley National Laboratory, 
Leibniz Institut f\"ur Astrophysik Potsdam (AIP),  
Max-Planck-Institut f\"ur Astronomie (MPIA Heidelberg), 
Max-Planck-Institut f\"ur Astrophysik (MPA Garching), 
Max-Planck-Institut f\"ur Extraterrestrische Physik (MPE), 
National Astronomical Observatory of China, New Mexico State University, 
New York University, University of Notre Dame, 
Observat\'orio Nacional / MCTI, The Ohio State University, 
Pennsylvania State University, Shanghai Astronomical Observatory, 
United Kingdom Participation Group,
Universidad Nacional Aut\'onoma de M\'exico, University of Arizona, 
University of Colorado Boulder, University of Oxford, University of Portsmouth, 
University of Utah, University of Virginia, University of Washington, University of Wisconsin, 
Vanderbilt University, and Yale University.

\begin{longrotatetable}
\begin{deluxetable*}{llcccccccccccc}
\centering
\scriptsize
\tablecaption{FGK dwarfs line-by-line Abundances} 
\label{tab:2} 
\tablewidth{700pt}
\tabletypesize{\scriptsize}
\tablehead{
Element	& Lambda 	&J12232820+&	J12265103+&	J12211561+&	J12285643+&	J12240572+&	J12270627+&	J12214901+&	J12204557+&	J12215616+&	J12234101+&	J12255195+\\
	&  (\AA{})	&2553400&	2616018&	2609140&	2632573&	2607430&	2650445&	2632568&	2545572&	2718342&	2658478&	2646359}
\startdata
FeI&	15194.5&	7.51&	7.42&	7.47&	7.51&	7.58&	7.55&	7.56&	... &	... &	... &	... \\
&	15207.5&	7.60&	7.57&	7.60&	7.66&	7.59&	7.57&	7.58&	7.60&	7.48&	7.48&	7.48\\
&	15224.5&	7.52&	7.49&	7.51&	7.46&	7.48&	7.46&	7.47&	... &	... &	... &	7.55\\
&	15239.9&	7.54&	7.54&	7.52&	7.48&	7.46&	7.51&	7.50&	... &	... &	... &	7.43\\
&	15245.0&	7.55&	7.53&	7.55&	7.58&	7.54&	7.53&	7.59&	7.65&	7.58&	7.55&	7.50\\
&	15294.5&	7.50&	7.40&	7.44&	7.48&	7.44&	7.43&	7.42&	7.50&	7.43&	7.43&	7.47\\
&	15301.4&	7.45&	7.53&	7.46&	7.46&	7.48&	7.45&	7.44&	7.44&	7.45&	7.50&	... \\
&	15343.8&	7.55&	7.52&	7.52&	7.48&	7.48&	7.48&	7.49&	... &	... &	... &	7.49\\
&	15395.7&	7.53&	7.48&	7.51&	7.52&	7.51&	7.50&	7.47&	7.49&	7.44&	7.47&	7.38\\
&	15490.3&	7.42&	7.38&	7.41&	7.40&	7.41&	7.43&	7.41&	7.47&	7.47&	7.49&	7.39\\
&	15496.4&	7.46&	7.38&	7.39&	7.40&	7.41&	7.41&	7.41&	7.36&	7.36&	7.38&	7.50\\
&	15501.1&	7.46&	7.39&	7.41&	7.39&	7.39&	7.41&	7.37&	7.41&	7.38&	7.39&	7.57\\
&	15531.8&	7.54&	7.49&	7.50&	7.51&	7.50&	7.48&	7.50&	7.48&	7.44&	7.50&	7.46\\
&	15534.2&	7.70&	7.59&	7.65&	7.68&	7.66&	7.60&	7.69&	7.60&	7.52&	7.57&	7.45\\
&	15537.9&	7.51&	7.47&	7.49&	7.43&	7.46&	7.42&	7.45&	7.39&	7.39&	7.46&	7.46\\
&	15588.0&	7.48&	7.45&	7.51&	7.46&	7.46&	7.45&	7.46&	7.47&	7.43&	7.45&	7.48\\
&	15648.5&	7.51&	7.41&	7.44&	7.45&	7.45&	7.47&	7.45&	7.47&	7.44&	7.46&	7.47\\
&	15662.0&	7.57&	7.51&	7.53&	7.55&	7.54&	7.52&	7.52&	7.55&	7.48&	7.48&	7.51\\
&	15677.0&	7.44&	7.43&	7.46&	7.43&	7.43&	7.44&	7.45&	7.45&	7.42&	7.47&	7.40\\
&	15686.2&	7.55&	7.53&	7.53&	7.49&	7.46&	7.47&	7.46&	7.49&	7.47&	7.51&	7.31\\
&	15692.8&	7.56&	7.50&	7.51&	7.49&	7.45&	7.44&	7.40&	7.39&	7.34&	7.40&	7.28\\
&	15904.4&	7.42&	7.36&	7.40&	7.37&	7.36&	7.35&	7.34&	7.32&	7.31&	7.31&	7.32\\
&	15906.1&	7.54&	7.50&	7.53&	7.54&	7.49&	7.45&	7.42&	7.39&	7.36&	7.28&	7.42\\
&	15911.5&	7.48&	7.45&	7.46&	7.46&	7.44&	7.45&	7.42&	7.33&	7.28&	7.32&	7.52\\
&	15913.0&	7.44&	7.41&	7.42&	7.37&	7.35&	7.32&	... &	... &	... &	... &	... \\
&	15921.0&	7.42&	7.40&	7.45&	7.42&	7.43&	7.41&	7.40&	7.47&	7.42&	7.42&	7.50\\
&	15964.9&	7.52&	7.49&	7.53&	7.55&	7.53&	7.53&	7.51&	7.55&	7.49&	7.52&	... \\
&	15980.8&	7.47&	7.44&	7.47&	7.47&	7.43&	7.45&	7.47&	7.55&	7.48&	7.50&	7.41\\
&	16006.9&	7.54&	7.51&	7.48&	7.42&	7.40&	7.39&	7.41&	7.46&	7.41&	7.41&	7.42\\
&	16009.6&	7.47&	7.43&	7.43&	7.44&	7.42&	7.42&	7.44&	7.50&	7.41&	7.42&	7.47\\
&	16038.0&	7.44&	7.41&	7.46&	7.42&	7.42&	7.44&	7.42&	7.51&	7.47&	7.47&	7.48\\
&	16040.7&	7.50&	7.48&	7.54&	7.50&	7.49&	7.51&	7.47&	7.56&	7.50&	7.48&	7.46\\
&	16043.0&	7.47&	7.44&	7.53&	7.45&	7.46&	7.47&	7.44&	7.56&	7.51&	7.46&	7.49\\
&	16076.0&	7.47&	7.44&	7.46&	7.43&	7.44&	7.43&	7.42&	7.52&	7.49&	7.49&	7.49\\
&	16088.9&	7.51&	7.51&	7.52&	7.50&	7.50&	7.51&	7.48&	7.40&	7.38&	7.49&	7.43\\
&	16102.4&	7.68&	7.65&	7.67&	7.72&	7.66&	7.59&	7.63&	7.51&	7.43&	7.43&	7.30\\
&	16116.0&	7.54&	7.53&	7.54&	7.55&	7.55&	7.55&	7.51&	7.42&	7.40&	7.30&	7.53\\
&	16126.0&	7.56&	7.52&	7.55&	7.58&	7.54&	7.54&	7.54&	7.57&	7.51&	7.53&	7.46\\
&	16153.2&	7.52&	7.51&	7.55&	7.53&	7.52&	7.48&	7.47&	7.52&	7.48&	7.46&	7.50\\
&	16165.0&	7.50&	7.48&	7.50&	7.50&	7.48&	7.48&	7.49&	7.58&	7.48&	7.50&	7.47\\
&	16175.0&	7.50&	7.48&	7.48&	7.44&	7.46&	7.45&	7.46&	7.50&	7.47&	7.47&	7.52\\
&	16178.0&	7.53&	7.52&	7.49&	7.50&	7.53&	7.53&	7.53&	... &	... &	... &	... \\
&	16180.0&	7.47&	7.45&	7.44&	7.44&	7.45&	7.45&	7.45&	7.51&	7.48&	7.52&	7.47\\
&	16185.8&	7.50&	7.47&	7.47&	7.42&	7.44&	7.44&	7.42&	7.44&	7.44&	7.47&	7.50\\
&	16195.0&	7.46&	7.43&	7.47&	7.46&	7.46&	7.47&	7.46&	7.49&	7.45&	7.50&	7.47\\
&	16198.1&	7.44&	7.44&	7.47&	7.51&	7.49&	7.49&	7.55&	7.63&	... &	... &	... \\
&	16204.0&	7.45&	7.41&	7.45&	7.46&	7.42&	7.45&	7.46&	7.44&	7.43&	7.47&	7.47\\
&	16207.6&	7.50&	7.45&	7.49&	7.46&	7.45&	7.45&	7.47&	7.51&	7.46&	7.47&	7.45\\
&	16213.5&	7.47&	7.44&	7.46&	7.44&	7.44&	7.43&	7.41&	7.42&	7.38&	7.45&	7.49\\
&	16232.0&	7.44&	7.41&	7.45&	7.44&	7.45&	7.45&	7.47&	7.56&	7.51&	7.49&	7.48\\
&	16236.0&	7.47&	7.47&	7.45&	7.43&	7.46&	7.43&	7.45&	7.51&	7.49&	7.48&	7.54\\
&	16246.0&	7.49&	7.46&	7.51&	7.46&	7.49&	7.49&	7.46&	7.51&	7.49&	7.54&	7.33\\
&	16292.8&	7.53&	7.51&	7.50&	7.49&	7.50&	7.49&	7.37&	7.26&	7.25&	7.33&	7.35\\
&	16316.3&	7.48&	7.45&	7.47&	7.51&	7.48&	7.49&	7.49&	7.64&	7.57&	7.57&	7.33\\
&	16324.5&	7.45&	7.40&	7.44&	7.44&	7.46&	7.43&	7.44&	7.42&	7.38&	7.33&	7.50\\
&	16331.5&	7.53&	7.49&	7.52&	7.49&	7.50&	7.49&	7.49&	7.54&	7.52&	7.50&	7.40\\
&	16394.4&	7.60&	7.57&	7.60&	7.61&	7.59&	7.56&	7.54&	7.41&	7.35&	7.40&	7.43\\
&	16398.1&	7.56&	7.53&	7.58&	7.58&	7.59&	7.53&	7.52&	... &	... &	... &	... \\
FeI&	16404.6&	7.60&	7.54&	7.62&	7.57&	7.59&	7.54&	7.49&	...&	...&	...&	...\\
&	16487.0&	7.43&	7.37&	7.40&	7.49&	7.42&	7.42&	7.41&	7.56&	7.47&	7.43&	7.29\\
&	16506.2&	7.52&	7.50&	7.49&	7.45&	7.47&	7.49&	7.46&	7.40&	7.37&	7.29&	7.49\\
&	16517.2&	7.47&	7.46&	7.46&	7.48&	7.45&	7.46&	7.48&	7.53&	7.47&	7.49&	7.49\\
&	16519.0&	7.42&	7.44&	7.45&	7.45&	7.47&	7.46&	7.41&	7.51&	7.47&	7.49&	...\\
&	16522.0&	7.48&	7.48&	7.47&	7.45&	7.47&	7.46&	7.43&	7.46&	7.43&	7.48&	7.48\\
&	16524.5&	7.50&	7.49&	7.51&	7.50&	7.48&	7.47&	7.49&	7.53&	7.47&	7.49&	7.49\\
&	16532.0&	7.52&	7.49&	7.53&	7.48&	7.49&	7.49&	7.49&	7.52&	7.49&	7.49&	7.49\\
&	16541.5&	7.47&	7.46&	7.46&	7.45&	7.48&	7.46&	7.43&	7.46&	7.44&	7.50&	7.50\\
&	16552.0&	7.46&	7.47&	7.46&	7.43&	7.45&	7.45&	7.45&	7.44&	7.42&	7.42&	7.42\\
&	16560.0&	7.45&	7.46&	7.47&	7.42&	7.47&	7.44&	7.42&	7.52&	7.48&	7.49&	7.49\\
&	16612.8&	7.47&	7.45&	7.45&	7.43&	7.44&	7.43&	7.39&	7.45&	7.43&	7.45&	7.45\\
&	16645.9&	7.48&	7.44&	7.46&	7.48&	7.49&	7.48&	7.47&	7.49&	7.46&	7.47&	7.47\\
&	16653.0&	7.43&	7.41&	7.42&	7.42&	7.44&	7.43&	7.41&	7.42&	7.40&	7.43&	7.43\\
&	16657.0&	7.57&	7.53&	7.67&	7.61&	7.69&	7.53&	7.65&	...&	...&	...&	...\\
&	16661.4&	7.59&	7.58&	7.62&	7.59&	7.60&	7.59&	7.59&	7.60&	7.56&	7.57&	7.57\\
&	16665.4&	7.52&	7.50&	7.54&	7.50&	7.51&	7.49&	7.52&	7.56&	7.54&	7.52&	7.52\\
&	&	&	&	&	&	&	&	&	&	&	&	\\
&	$\langle$A(FeI)$\rangle$&	7.50&	7.47&	7.49&	7.48&	7.48&	7.47&	7.47&	7.49&	7.44&	7.46&	7.45\\
&	$\langle$[FeI/H]$\rangle$&	0.05&	0.02&	0.04&	0.03&	0.03&	0.02&	0.02&	0.04&   -0.01&	0.01&	0.00\\
&	$\sigma$&	0.06&	0.06&	0.06&	0.07&	0.06&	0.05&	0.06&	0.08&	0.06&	0.07&	0.07\\
\label{FGK_abu}
\enddata
\end{deluxetable*}
\end{longrotatetable}


\facility {Sloan}

\software{Turbospectrum (\citealt{AlvarezPLez1998}, \citealt{Plez2012}), MARCS (\citealt{Gustafsson2008}), Bacchus (\citealt{Masseron2016}), Matplotlib (\citealt{matplotlib}), Numpy (\citealt{numpy}).}


{}



\end{document}